\documentclass[preprint2,numberedappendix]{emulateapj}

\newcommand{\Msun}{M\ensuremath{_{\odot}}}
\newcommand{\Msunyr}{M\ensuremath{_{\odot}}yr\ensuremath{^{-1}}}
\newcommand{\ergscm}{erg\,s\ensuremath{^{-1}}\,cm\ensuremath{^{-2}}}

\newcommand{\kms}{{\rm km\,s}$^{-1}$}

\newcommand{\Ha}{H$\alpha$}
\newcommand{\Hb}{H$\beta$}
\newcommand{\Nii}{[N{\sc ii}]}
\newcommand{\Oiii}{[O{\sc iii}]}
\newcommand{\Oii}{[O{\sc ii}]}

\def\lsim{\mathrel{\rlap{\lower4pt\hbox{\hskip1pt$\sim$}}
    \raise1pt\hbox{$<$}}}                
\def\gsim{\mathrel{\rlap{\lower4pt\hbox{\hskip1pt$\sim$}}
    \raise1pt\hbox{$>$}}}                

\usepackage[small,normal,bf,up]{caption}
\renewcommand{\captionfont}{\small}

\usepackage{natbib,natbibspacing}
\usepackage{amssymb,amsmath}

\shorttitle{Metallicity Gradients at $z\simeq2$}

\shortauthors{Jones, Ellis, Richard, \& Jullo}

\begin{document}


\title[Metallicity Gradients at $z\simeq2$]{The Origin and Evolution of Metallicity Gradients: \\
Probing the Mode of Mass Assembly at $z\simeq2$}

\author{Tucker Jones\altaffilmark{1}, Richard S. Ellis\altaffilmark{1}, Johan Richard\altaffilmark{2}, Eric Jullo\altaffilmark{3}
}

\altaffiltext{1}{Astronomy Department, California Institute of Technology, MC249-17, Pasadena, CA 91125, USA}
\altaffiltext{2}{CRAL, Observatoire de Lyon, Universit\'e Lyon 1, 9 Avenue Ch. Andr\'e, 69561 Saint Genis Laval Cedex, France}
\altaffiltext{3}{Laboratoire d'Astrophysique de Marseille, Universit\'e d'Aix-Marseille \& CNRS, UMR7326, 38 rue F. Joliot-Curie, 13388 Marseille Cedex 13, France}



\begin{abstract}

We present and discuss measurements of the gas-phase metallicity gradient in gravitationally lensed galaxies at $z=2.0-2.4$ based on adaptive optics-assisted imaging spectroscopy with the Keck II telescope. Through deep exposures we have secured high signal to noise data for four galaxies with well-understood kinematic properties.
Three galaxies with well-ordered rotation reveal metallicity gradients in the sense of having lower gas-phase metallicities at larger galactocentric 
radii. Two of these display gradients much steeper than found locally, while a third has one similar to that seen in local disk galaxies. The fourth galaxy exhibits complex 
kinematics indicative of  an ongoing merger and reveals an ``inverted'' gradient with lower metallicity in the central regions. By comparing our sample to similar data in the 
literature for lower redshift galaxies, we determine that, on average, metallicity gradients must flatten by a factor of $2.6\pm0.9$ between $z=2.2$ and the present epoch. 
This factor is in rough agreement with the size growth of massive galaxies suggesting that inside-out growth can account for the evolution of metallicity gradients. Since
the addition of our new data provides the first indication of a coherent picture of this evolution, we develop a simple model of chemical evolution to explain the collective data. We find 
that metallicity gradients and their evolution can be explained by the inward radial migration of gas together with a radial variation in the mass loading factor governing the
ratio of outflowing gas to the local star formation rate. Average mass loading factors of $\lsim2$ are inferred from our model in good agreement with direct measurements of outflowing gas in $z\simeq2$ galaxies.

\end{abstract}

\section{Introduction}

Metallicity is one of the most fundamental properties of a galaxy. Gas-phase metallicity (hereafter referred to simply as metallicity) is governed by the cumulative history of baryonic assembly: gas accretion, star formation, and gas outflow. Metallicity is therefore a tracer of the processes responsible for galaxy formation and evolution. The vast database of information made available by the Sloan Digitized Sky Survey (SDSS) has revealed that metallicity is correlated with galaxy stellar mass and anticorrelated with star formation rate \citep{Tremonti04, Mannucci10, Lopez10}. The dependence on stellar mass is explained as the effect of gravity: more massive galaxies lose a smaller fraction of their metal-enriched gas in outflows and are more easily able to re-accrete lost metals due to their stronger gravitational potential \citep{Tremonti04}. The dependence on star formation rate is primarily an effect of accretion: increased accretion of metal-poor gas drives higher star formation rates and lowers the overall metallicity (e.g. \citealt{Dave11}).

The spatial distribution of metallicity and its evolution with time is sensitive to the assembly history of galaxies. Careful measurements in the local universe show that all disk galaxies exhibit negative radial metallicity gradients, with lower metallicity at larger galactocentric radius (e.g. \citealt{Vila-Costas92, vanZee98, Considere00, Rupke10}). In many cases the gradient flattens at large radius indicating efficient radial mixing, possibly induced by interactions or cycling between the disk and halo in a ``galactic fountain'' process (e.g. \citealt{Werk11, Bresolin12}). More attention has been given recently to measurements of the time evolution of metal gradients and the implications for galaxy formation. Models of inside-out disk growth predict that gradients should be initially steep and become flatter at later times (e.g. \citealt{Prantzos00, Magrini07, Fu09, Marcon-Uchida10}). In contrast, numerical simulations show that merger-driven growth will rapidly flatten existing metallicity gradients, which then become steeper with time \citep{Rupke10a}. This is supported by nearby observations which reveal flatter gradients in interacting galaxies compared to an isolated control sample \citep{Rupke10}.

Several groups have now begun to consider the time evolution of radial metallicity gradients through observations. One approach is to examine the situation in detail within the Milky Way. \cite{Maciel03} have inferred that the metallicity gradient must have been steeper in the past by analyzing the properties of stellar systems with different characteristic ages. However, this method is only feasible in a small number of local galaxies.
The alternative approach, adopted here, is to measure the gradient in resolved data for galaxies seen at high redshift and to then compare these measures with those at lower redshift. Although a challenging task, this was first attempted in a previous paper where we showed the metallicity gradient in a gravitationally lensed galaxy at $z=2$ was significantly steeper than in local disk galaxies \citep{Jones10b}. Another lensed galaxy at $z=1.5$ yielded similar results \citep{Yuan11}, further suggesting an inside-out mode of galaxy growth. However, some studies at high redshift have suggested that many massive galaxies may have "inverted" (positive) gradients with lower metallicity in the central regions \citep{Cresci10, Queyrel12}. This would imply a radically different mode of growth. One issue with this approach is the technical difficulty of measuring a reliable gradient in seeing-limited data which offers a relatively poor spatial resolution of $\simeq 4$ kpc FWHM, approximately twice the typical half-light radius of an $L_*$ galaxy at $z=2-3$ \citep{Bouwens04}. Such measurements cannot provide good spatial sampling except for the very largest sources. Clearly it is advantageous to use adaptive optics as well as the angular magnification afforded by gravitational lensing.

A further issue concerns the fact that some high redshift metal gradients reported in the literature are measured from a limited set of emission lines which may be affected by shocks and/or AGN (e.g. \Nii/\Ha, \Oiii/\Hb). Shocks can easily produce a signature indicative of an inverted metallicity gradient in the \Nii/\Ha\, line ratio (e.g. \citealt{Westmoquette12}), while AGN can mimic an inverted gradient in \Oiii/\Hb\, and a strong negative gradient in \Nii/\Ha\, (e.g. \citealt{Wright10}). Hence multiple emission line diagnostics are required to distinguish variations in metallicity from the effects of shocks and AGN following methods such as those of \cite{Baldwin81}. While observations of multiple emission lines required for such diagnostics are challenging to obtain, they are necessary for robust metallicity gradient measurements.

This paper is concerned with increasing the number of galaxies at high redshift with reliable high-resolution metallicity measurements. Gravitationally lensed galaxies are ideally suited to this purpose, as the strong magnification provides both an increased flux and higher source plane resolution (e.g. \citealt{Jones10a,Jones10b}). In this paper we present spatially resolved metallicity measurements of four lensed galaxies at $z=2-2.4$ including one previously studied by \cite{Jones10b}. For the first time this permits us to consider the overall evolution of the metallicity gradient by utilizing lower redshift data. In \S2 we describe the source selection, gravitational lens models, observations, and data reduction. In \S3 we discuss the kinematic properties of each source which provides valuable insight into variations we see within our sample. In \S4 we present resolved metallicity measurements. In \S5 we construct and discuss a simple chemical evolution model to describe the origin of metallicity gradients in the context of the growing body of data we assemble for comparison purposes. The details of this model are given in an Appendix. In \S6 we discuss the results in the context of our chemical evolution model. Finally, in \S7 we summarize the main conclusions and implications of the results.

Throughout this paper we adopt a $\Lambda$CDM cosmology with $H_0$= 70 km s$^{-1}$ Mpc$^{-1}$, $\Omega_M$=0.30 and $\Omega_\Lambda$=0.70. At $z$=2.2, 0.1 arcsec corresponds to 830 pc and the age of the universe was 2.9 Gyr. All magnitudes are in the AB system.

\section{Observations and Data Analysis}

\begin{table*}
\begin{tabular}{lccccccccc}
Name   & $z$  &  Coordinates  & Dates  &  Filter  & Lines  & $T_{int}$ (s) &  FWHM  &  FWHM           & $\mu$ \\
       &      &               &        &          &        &               &  (PSF) &  (source plane) &       \\
\hline
J0744  & 2.21 &  07:44:47.9 +39:27:26  &  27 Nov 2008                  &  Kn2  &  \Ha, \Nii  &  14400 &  0\farcs11  &  $0.3\times0.8$ kpc  & $16 \pm 3$     \\
       &      &                        &  20 Feb 2011                  &  Hn2  &  \Oiii, \Hb &  10800 &  0\farcs08  &  $0.3\times0.7$ kpc  \\
J1038  & 2.20 &  10:38:41.8 +48:49:19  &  20 Feb 2011, 12 Mar 2011     &  Kn2  &  \Ha, \Nii  &  4800  &  0\farcs14  &  $0.4\times1.6$ kpc  & $8.4 \pm 0.7$  \\
       &      &                        &  20 Feb 2011                  &  Hn2  &  \Oiii, \Hb &  1200  &  0\farcs14  &  $0.3\times1.7$ kpc  \\
J1148  & 2.38 &  11:48:33.3 +19:29:59  &  20 Feb 2011                  &  Kc4  &  \Ha, \Nii  &  7200  &  0\farcs11  &  $0.6\times0.9$ kpc  & $10.3 \pm 5.0$ \\
       &      &                        &  20 Feb 2011, 12 Mar 2011     &  Hbb  &  \Oiii, \Hb &  3600  &  0\farcs08  &  $0.6\times0.9$ kpc  \\
J1206  & 2.00 &  12:06:01.7 +51:42:30  &  19 May 2010                  &  Kn1  &  \Ha, \Nii  &  3600  &  0\farcs18  &  $0.5\times3.0$ kpc  & $13.1 \pm 0.7$ \\
       &      &                        &  19 May 2010                  &  Hn1  &  \Oiii      &  1800  &  0\farcs33  &  $0.6\times3.5$ kpc  \\
\end{tabular}
\caption{\label{tab:obs} Log of observations.}
\end{table*}

\subsection{Source Selection}

Our sources were selected as ideal targets for resolved metallicity measurement based on their redshift, nebular emission line intensity, magnification, and the presence of a suitable star for tip/tilt correction. Our lensed sources were selected to lie at $z=2.0-2.4$ such that the diagnostic emission lines \Ha, \Hb, \Nii, and \Oiii\, can be observed in the H and K atmospheric windows where the Keck II adaptive optics (AO) system provides the best Strehl ratio. The exception is SDSS J1206 for which \Hb\ lies in a telluric absorption band and is inaccessible with OSIRIS. Prior to AO-assisted observation, we secured a near infrared spectrum of each source to ensure that the nebular emission lines are sufficiently bright and relatively free of contamination from OH sky lines \citep{Richard11}. We require that each gravitational lens system has an accurate mass model constrained by the positions of multiply-imaged background sources with known redshift, essential for the source plane reconstruction discussed in \S\ref{sec:lens}. Finally, we require a suitably bright and nearby star to provide tip/tilt correction for the adaptive optics system. The observational details of each galaxy are given in Table~\ref{tab:obs}. Below we provide a brief overview of the sources described in this paper. 

{\em MACS J0744+3927} (hereafter J0744) is a galaxy cluster lensing system discovered by the MAssive Cluster Survey \citep{Ebeling01}. The $z=2.21$ arc was identified in followup optical spectroscopy of the cluster and we secured the near-infrared spectrum as part of our screening program of bright $z\gsim2$ arcs \citep{Richard11}. We previously observed the \Ha\ emission line with OSIRIS and described the source kinematics in \cite{Jones10a}.

{\em SDSS J1038+4849} (hereafter J1038) is a group-scale lensing system discovered in the CASSOWARY survey for wide-separation gravitational lenses in SDSS imaging \citep{Belokurov09}. We obtained a near-infrared spectrum of the $z=2.20$ arc using NIRSPEC on Keck which showed bright nebular emission lines suitable for resolved spectroscopy with OSIRIS. We show in \S\ref{sec:kinematics} that the source is a merger of at least two systems with stellar mass ratio of (6$\pm$3):1. Therefore in much of the analysis we consider the two components separately. The components are divided into the region with y-axis $>-1$ kpc in Figure~\ref{fig:morphology} (``North'' region) and y-axis $<-1$ kpc (``South''). The northern region is brighter in optical WFPC2 imaging and the southern region is brighter in IRAC.

{\em SDSS J1148+1930} (hereafter J1148) is comprised of a background source at $z=2.38$ lensed into a nearly complete Einstein ring by a massive galaxy at $z=0.44$. The system was discovered in the SDSS by \cite{Belokurov07}, and near-infrared longslit spectroscopy of the arc is reported in \cite{Hainline09}.

{\em SDSS J1206+5142} (hereafter J1206) is a group-scale lensing system discovered in the SDSS by \cite{Lin09}. We previously obtained resolved metallicity measurements of this source from observations of \Ha, \Nii, and \Oiii\, with OSIRIS \citep{Jones10b}. J1206 has an unusually steep radial metallicity gradient compared to local galaxies, motivating high-resolution studies of additional sources presented in this paper. We include this source in the analysis and discussion of the present paper for completeness.

\begin{figure*}
\hspace{-.1\textwidth}
\includegraphics[width=0.7\textwidth]{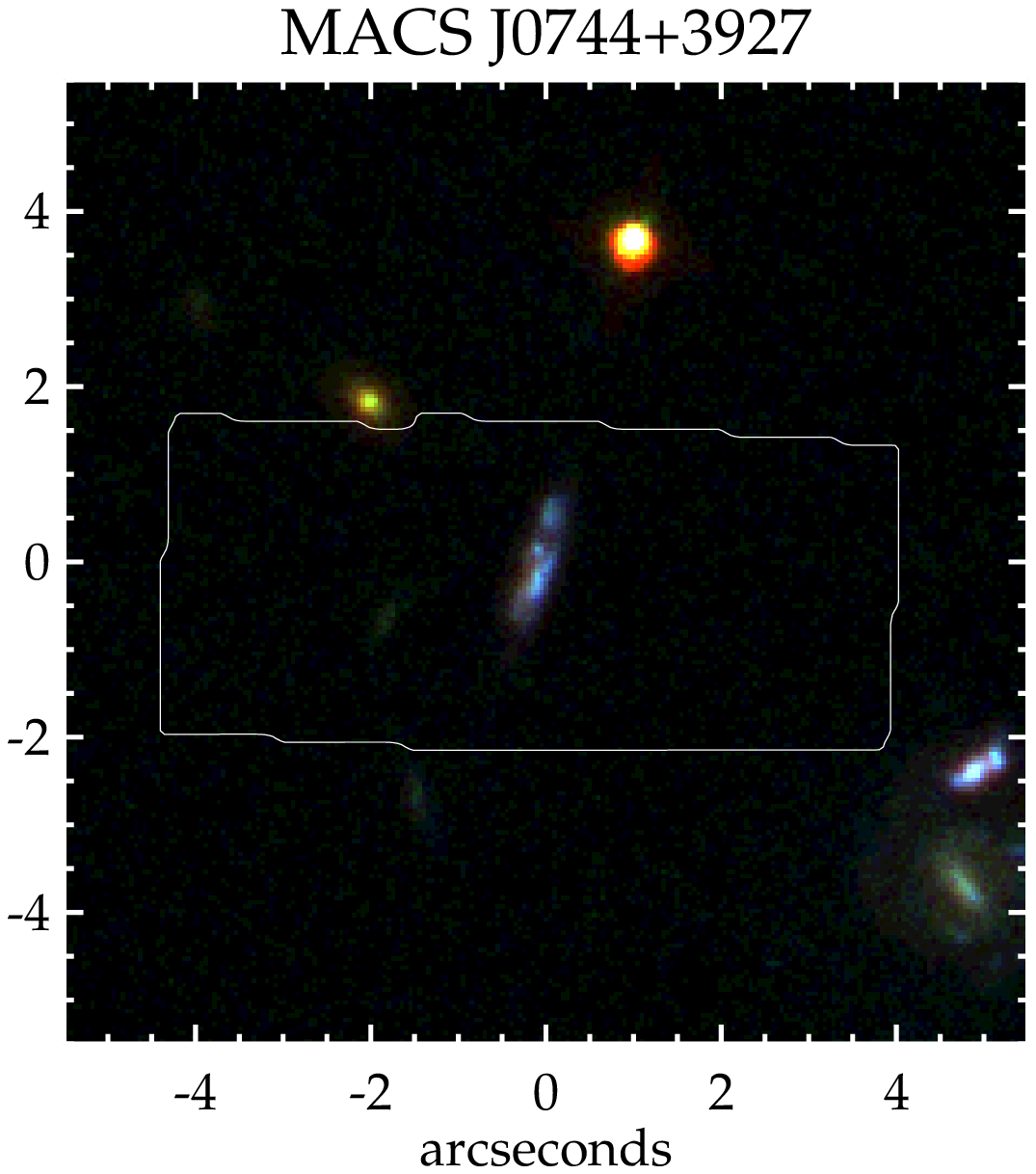}
\hspace{-.2\textwidth}
\includegraphics[width=0.7\textwidth]{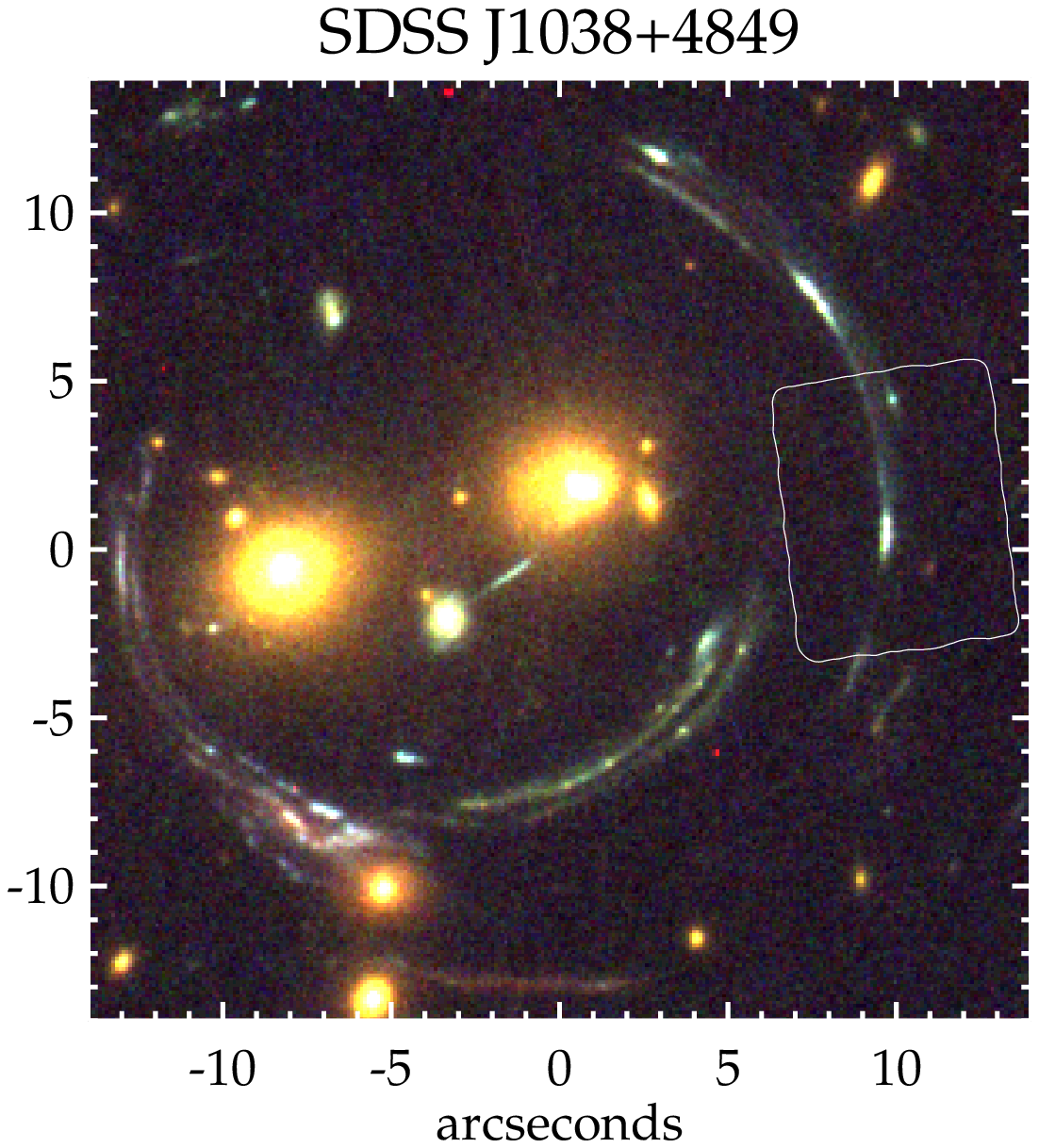} \\

\hspace{-.1\textwidth}
\includegraphics[width=0.7\textwidth]{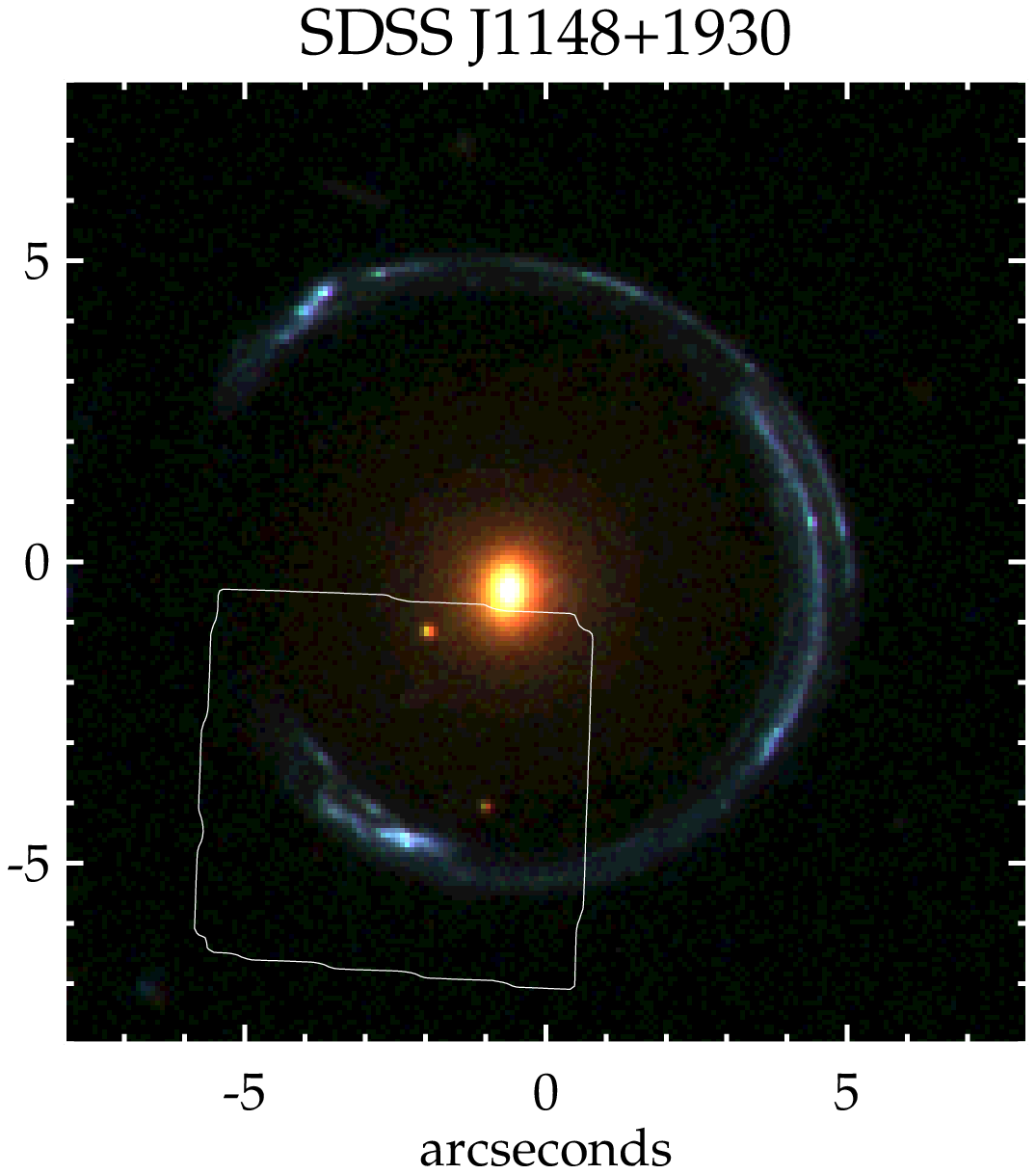}
\hspace{-.2\textwidth}
\includegraphics[width=0.7\textwidth]{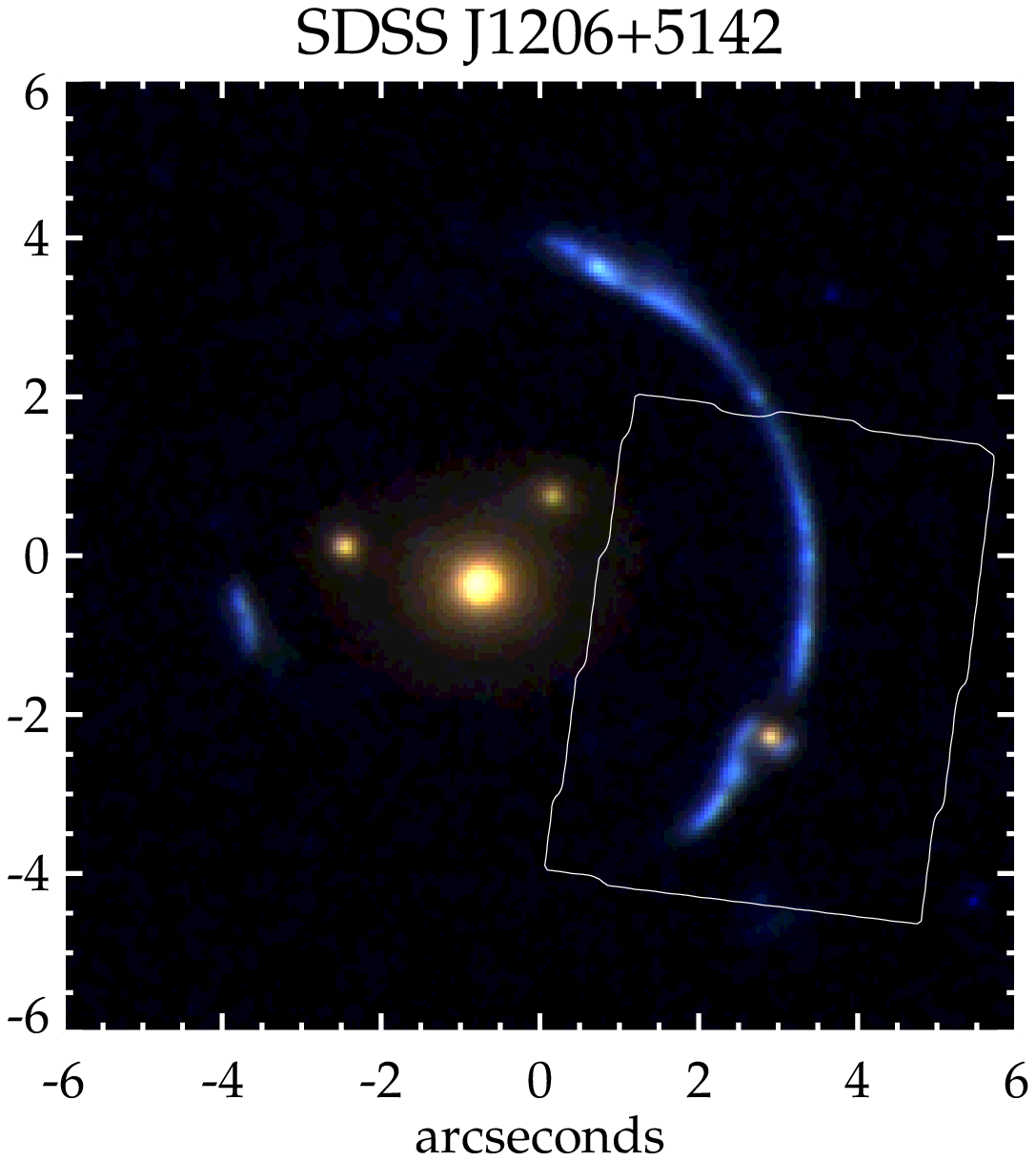}
\caption{\label{fig:sample} 
Color {\sl HST} images of the sample. The white box in each image shows the OSIRIS field of view used to observe \Ha\, and \Nii\, emission lines. R/G/B channels of each image are respectively ACS F814W/F555W/WFC3 F390W for J0744, WFPC2 F814W/F606W/F450W for J1038 and J1206, and WFC3 F814W/F606W/F475W for J1148.
}
\end{figure*}

\subsection{Gravitational Lens Models}\label{sec:lens}

Accurate models of the foreground mass distribution are essential for correcting the gravitational lensing distortion and reproducing the source-plane morphology of background galaxies. Here we briefly describe the adopted procedure. We use the {\sc Lenstool} program \citep{Kneib93, Jullo07} to parameterize the mass distribution of the lens, using the positions of multiply-imaged background sources as constraints. The mass models used for J0744 and J1206 are described in Limousin et al.\~(in preparation) and \cite{Jones10b} respectively.
We constructed lensing mass models for the two other systems based on available {\em HST} imaging and multiple images with spectroscopic redshifts. In the case of J1038, we use the two triple systems at $z=0.9657$ and $z=2.198$ \citep{Belokurov09} to constrain a double pseudo-isothermal elliptical (dPIE) dark matter profile (including core $r_{\rm core}$ and cut $r_{\rm cut}$ radii, see \citealt{Jullo09} for more details). Cut radii are not constrained by strong lensing and so we use a fixed value $r_{\rm cut} = 800$ kpc typical of galaxy groups.
For J1148 we use the system of 4 images at $z=2.38$ forming the `Horseshoe' as constraints for a dPIE mass distribution, following the modelling of \cite{Dye08}. In both lens systems, individual group galaxies detected in the optical bands were added as low-scale dPIE mass components perturbing the model, a procedure similar to the one used for J1206 in \cite{Jones10b}. The best fit parameters of the dPIE potentials for these previously unpublished cases are summarized in Table~\ref{tab:lensmodels}.

\begin{figure*}
\centerline{\includegraphics[width=0.85\textwidth]{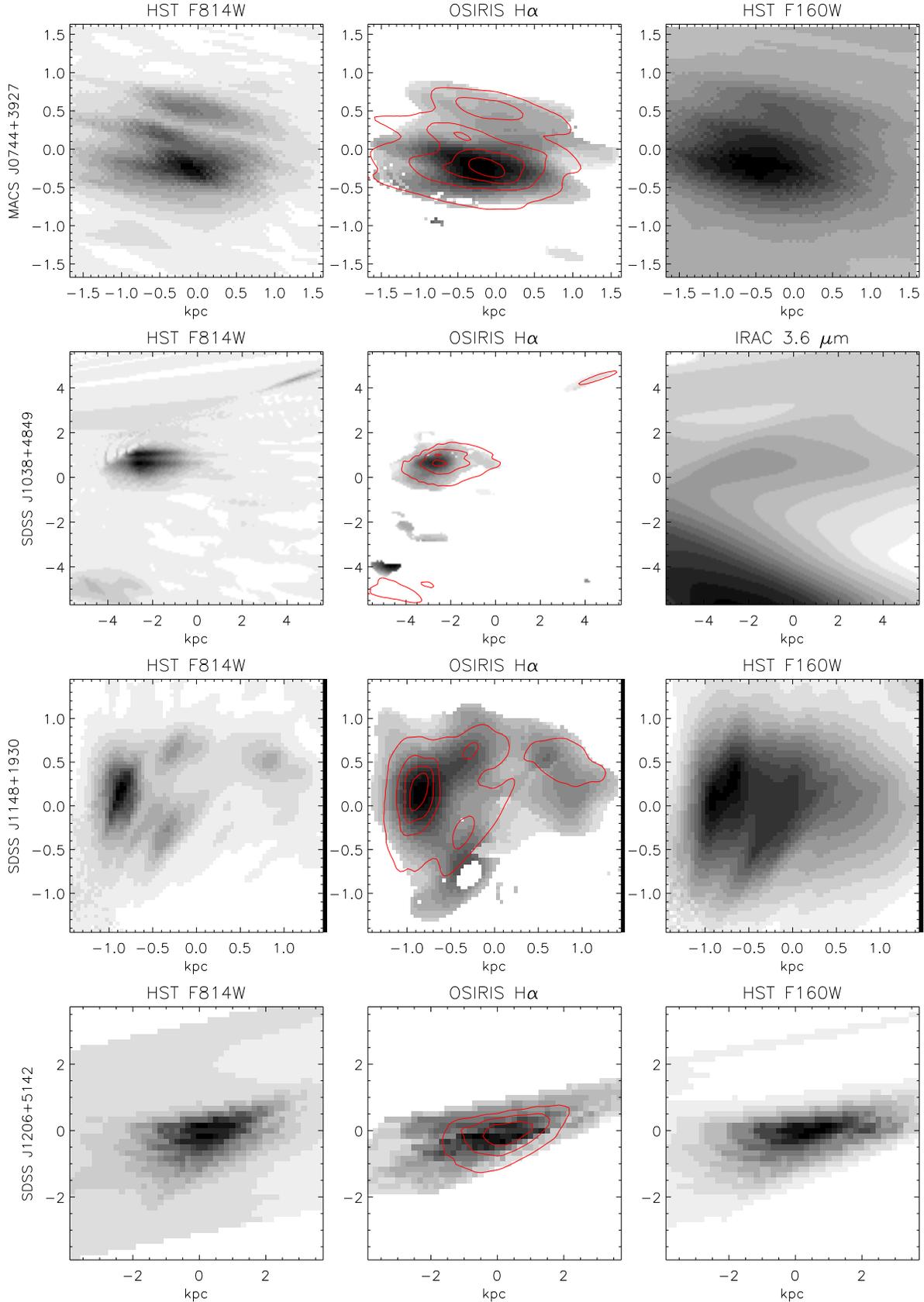}}
\caption{\label{fig:morphology} 
Source plane reconstruction of the lensed galaxies. Each row shows the morphology of one galaxy in (from left to right) rest-UV continuum, \Ha\, emission, and rest-optical continuum. No rest-optical continuum image is available for SDSS J1038 so we show the reconstructed IRAC 3.6$\mu$m image.
In all sources the ionized gas morphology traced by \Ha\, is similar to that of the rest-UV continuum, shown as contours of {\sl HST/ACS} F814W intensity on the \Ha\, maps.
}
\end{figure*}

{\sc Lenstool} also determines the transformation from image plane to source plane position, and we use this mapping to reconstruct all observations onto a uniform grid in the source plane. Intrinsic source-plane morphologies of each galaxy are shown in Figure~\ref{fig:morphology}. Lensing magnification is calculated as the ratio of \Ha\, flux in the image plane to that in the source plane, or equivalently the ratio of area subtended by \Ha\, emission. In Table~\ref{tab:obs} we give the typical magnification and $1\sigma$ lens model uncertainty for the regions observed with OSIRIS. In the case of J1206, multiple images are covered with OSIRIS such that the total magnification of images within the field of view is $\simeq 20$.

\begin{table}
\begin{tabular}{lccccccc}
Parameter   &  J1038         &  J1148  \\
\hline
$X_C\,^a$     &  -2.4$\pm$0.7  &  -0.11$\pm$0.11  \\
$Y_C\,^a$     &  -2.6$\pm$0.5  &  0.47$\pm$0.13   \\
$e^b$       &  0.76$\pm$0.03 &  0.00$\pm$0.02   \\
$\theta^c$  &  31$\pm$1      &  134$\pm$12      \\
$\sigma^d$  &  576$\pm$12    &  654$\pm$78      \\
$r_{core}\, ^e$ &  34.1$\pm$2.9  &  30.9$\pm$10.2   \\
$r_{cut}\, ^e$  &  [800]         &  [800]         \\
\end{tabular}

$^a$ Position from the brightest galaxy of the group (arcseconds). \\
$^b$ Ellipticity. \\
$^c$ Position angle (degrees). \\
$^d$ Velocity dispersion (\kms). \\
$^e$ Core and cut radii (kpc). \\
\caption{\label{tab:lensmodels} Best fit parameters for the previously unpublished lensing potentials (see text).
}
\end{table}

\subsection{Integral Field Spectroscopy}

Spectroscopic observations of the targets in Table~\ref{tab:obs} were taken with the OH-Suppressing Infra-Red Imaging Spectrograph (OSIRIS; \citealt{Larkin06}) in conjunction with the laser guide star assisted adaptive optics (AO) system \citep{Wizinowich06} on the Keck II telescope. These were carried out during four separate observing runs with seeing ranging from 0\farcs7 to 1\farcs5 and clear conditions as summarized in Table~\ref{tab:obs}. 
A suitably bright star ($R < 17$) within 55 arcseconds of each target was used for tip-tilt correction. We used the 100 milliarcsecond pixel scale in all observations which provided a field of view between 1\farcs6$\times$6\farcs4 and 4\farcs5$\times$6\farcs4 depending on the filter. Observations of each target were done with an AB observing sequence, dithering by $\sim 2-3$ arcseconds to keep the target within the integral field unit. The fields of view for \Ha\, and \Nii\, are shown for reference in Figure~\ref{fig:sample}. Fields of view for \Oiii\, and \Hb\, are essentially identical except for J1038, where only the northern region was observed.

Our data reduction methods are essentially identical to those described in \cite{Jones10b}. We used the OSIRIS Data Reduction Pipeline (ODRP; \citealt{Larkin06}) to perform dark and bias subtraction, cosmic ray rejection, wavelength calibration, and to assemble the 3D data cubes. Sky subtraction was done with the IDL code described in \cite{Davies07} using temporally adjacent exposures as sky reference frames. Final data cubes were combined using a $\sigma$-clipped mean. Flux calibration for J1206 and the K-band observations of J0744 are described in \cite{Jones10b} and \cite{Jones10a} respectively. For the remaining observations, we follow the same method as in \cite{Jones10a} using observations of the tip/tilt reference stars to calibrate the absolute flux. We additionally check the flux calibration with observations of the UKIRT infrared standard star FS 26 taken at a different time on the same nights. Flux calibrations derived from FS 26 agree to within 15\% suggesting the systematic uncertainty in flux calibration is $\lsim 15$\%.

\subsubsection{Extinction and Star Formation Rate}

In this subsection we determine the extinction and star formation rate of each galaxy from their integrated spectra.
We determine the total flux of nebular emission lines by summing all pixels within 250 \kms\, of the systemic redshift and within $\sim 0\farcs5$ of regions where the arc is detected in {\em HST} imaging. The results are shown in Table~\ref{tab:flux}. Balmer line ratios \Ha/\Hb\, are used to determine the dust extinction assuming a \cite{Calzetti00} reddening curve, such that
\begin{equation}
\mathrm{E(B-V)} = 1.695 \, \log \frac{\mathrm{H}\alpha / \mathrm{H}\beta}{2.86}
\end{equation}
where \Ha/\Hb\,$=2.86$ is the intrinsic line ratio for case B recombination. The extinction in V band and in \Ha\, are then given by
\begin{equation}
\mathrm{A_V} = \mathrm{R_V \, E(B-V)} = 4.05 \, \mathrm{E(B-V)}
\end{equation}
\begin{equation}
\mathrm{A_{H\alpha}} = 3.33 \, \mathrm{E(B-V)}
\end{equation}
for the \cite{Calzetti00} reddening curve. Star formation rates are computed from total \Ha\, flux corrected for extinction and lensing magnification (Table~\ref{tab:obs}), with the conversion factor from \cite{Kennicutt98} for a Salpeter IMF. The resulting E(B-V), A$_{\mathrm{V}}$, SFR, and 1$\sigma$ uncertainties for each source are listed in Table~\ref{tab:mass}. We note that the lower bound on SFR is required to be greater than or equal to the \Ha-derived SFR assuming no extinction. \Hb\, is not observed for the southern region of J1038 and so we give only the extinction-free SFR as a lower limit. For J1206 we use the Balmer line ratios reported by \cite{Hainline09} to derive E(B-V)\,$=0.30 \pm 0.12$ and apply this extinction correction to the \Ha\, flux measured with OSIRIS.

\begin{table}
\begin{tabular}{lccccc}
Name  &  \Ha  &  \Nii  &  \Oiii$_{5007}$  &  \Oiii$_{4959}$  &  \Hb  \\
\hline
J0744  &  $20\pm2$  &  $9\pm2$  &  $12\pm2$  &  $6\pm2$   &  $6\pm2$   \\
J1038 (North)  &  $66\pm5$  &  $5\pm5$  &  $100\pm5$ &  $42\pm8$  &  $19\pm7$  \\
J1038 (South)  &  $71\pm6$ &  $0\pm7$  \\
J1148  &  $67\pm3$  &  $7\pm3$  &  $41\pm2$  &  $8\pm2$   &  $10\pm3$  \\
J1206  &  $236\pm7$ &  $51\pm7$ & $230\pm11$ & $108\pm14$ &  \\
\end{tabular}
\caption{\label{tab:flux} Total observed emission line fluxes. All values are in units of $10^{-17}$ \ergscm. Uncertainties are $1\sigma$ determined from noise in the integrated spectra and do not include systematic errors in absolute flux calibration, which are typically $\lsim 15$\%. For J1038 we report the northern and southern regions separately; the southern region was not observed in \Oiii\, and \Hb.}
\end{table}

\subsection{Photometry and Stellar Mass}

For an appropriate comparison with lower redshift data, it will be helpful to estimate the masses of our target galaxies.
We have determined stellar masses for each galaxy from multi-wavelength photometry and stellar population synthesis modelling. All of our sources have {\it HST} imaging in multiple optical filters (e.g. Figure~\ref{fig:sample}) as well as {\it Spitzer} IRAC channels 1 and 2. All galaxies except J1038 additionally have high resolution near-infrared imaging from {\it HST} WFC3/IR. We determine the photometry in matched apertures and then scale to the total flux. All data are first smoothed to match the point spread function of IRAC channel 1, which has the lowest spatial resolution. We then measure the flux in each filter within an aperture selected to encompass the majority of the arc while avoiding contamination from nearby sources. The total flux is determined from high-resolution optical {\it HST} images and all aperture fluxes are scaled by the same factor to obtain the total flux. Finally, near-infrared fluxes are corrected for nebular emission by subtracting the total line fluxes (Table~\ref{tab:flux}).
We fit the photometric data with stellar population models using the code {\sc FAST} \citep{Kriek09} with an exponentially declining star formation history, Salpeter initial mass function, and \cite{Calzetti00} reddening law. Redshifts are fixed to the spectroscopic values and metallicity is allowed to vary within the spectroscopic values for each arc (e.g. Table~\ref{tab:gradients2}). We allow a range of star formation timescales $10^7-10^{10}$ yr. Ages are restricted to be $>50$ Myr (approximately equal to the dynamical timescale for these sources; see \S\ref{sec:kinematics}) and less than the age of the universe ($\simeq 3$ Gyr at $z=2-2.4$). Dust extinction is allowed to vary within the $\pm1\sigma$ range of A$_{\mathrm{V}}$ in Table~\ref{tab:mass} and can be as low as half the allowed minimum value, since stellar continuum is often inferred to have a factor of $\simeq2$ lower A$_{\mathrm{V}}$ than H{\sc ii} regions (e.g. \citealt{Newman12}). In all cases we require A$_{\mathrm{V}} \geq 0$, and for the southern region of J1038 we assume $A_V\leq4$. Finally, we consider only star formation histories which produce a SFR consistent to within $1\sigma$ of the \Ha-derived values. The best fit stellar mass and 68\% confidence levels within these constraints are given in Table~\ref{tab:mass}.

Correcting photometric measurements for contamination by nebular emission lines may be important for accurately determining stellar masses of star forming galaxies. Our sources have emission lines with rest-frame equivalent widths $W_0 \simeq 50$ \AA\, for \Oiii$_{5007}$ and \Ha. At redshift $z=2$ this corresponds to roughly 10\% of the flux measured in a broad-band filter. Indeed, \Oiii\, and \Hb\, contribute $\sim 20$\% of total H-band flux measured for the lensed sample. To examine the effect on derived stellar mass, we apply the same stellar population synthesis method (and the same constraints) without correcting for nebular emission lines. This results in best-fit stellar masses which are higher by $+0.21$ dex for J0744, $+0.002$ dex for J1148, and $+0.05$ dex for J1206. In general the correction is not large and is consistent in all cases within the uncertainties, however in the case of J0744 the stellar mass is overestimated by 60\% if nebular emission is not accounted for. In summary, ignoring the nebular emission contribution would bias the stellar masses to somewhat higher values, but not significantly affect any results in this paper.

\begin{table*}
\begin{tabular}{lccccc}
				&	J0744		&	J1038 North	&	J1038 South	&	J1148		&	J1206		\\
\hline
$\log M_*$ (\Msun)		&  $9.4^{+0.4}_{-0.1}$	&  $9.1^{+0.2}_{-0.1}$	&  $9.9^{+0.2}_{-0.3}$	&  $9.9^{+0.2}_{-0.3}$	&  $10.1^{+0.2}_{-0.2}$	\\
E(B-V)				&  $0.13 \pm 0.27$	&  $0.17 \pm 0.32$	&			&  $0.73 \pm 0.26$	&  $0.30 \pm 0.12$	\\
A$_{\mathrm{V}}$  		&  $0.53 \pm 1.20$	&  $0.67 \pm 1.30$	&			&  $2.94 \pm 1.05$	&  $1.22 \pm 0.47$	\\
SFR (\Msunyr)			&  $5.4^{+4.9}_{-1.8}$	&  $38^{+37}_{-15}$	&  $>24$		&  $210^{+167}_{-167}$	&  $68^{+44}_{-24}$	\\
$M_{\mathrm{gas}}$ ($10^9$ \Msun)  &  $1.8^{+1.1}_{-0.5}$  &			&			&  $27^{+14}_{-18}$	&  $14^{+3}_{-4}$	\\
$f_{\mathrm{gas}}$       	&  $0.42^{+0.40}_{-0.08}$  &			&		     &  $0.77^{+0.14}_{-0.15}$	&  $0.52^{+0.25}_{-0.11}$	\\
$\log M_{\mathrm{halo}}$ (\Msun) &  11.6		&  11.4			&  11.8			&  11.9			&  12.0			\\
D (kpc)				&  2.4			&  13.3			&			&  2.7			&  3.4			\\
$\Delta V$ (\kms)		&  $252 \pm 33$		&  $160 \pm 10$		&			&  $148 \pm 3$		&  $159 \pm 38$		\\
$\sigma$ (\kms)			&  $89 \pm 26$		&  $82 \pm 22$		&			&  $90 \pm 33$		&  $104 \pm 37$		\\
$M_{dyn}$ ($10^{10}$ \Msun)	&  1.1			&			&			&  1.3			&  2.1			\\
$Q / \sin{i}$			&  0.5			&			&			&  0.3			&  0.3			\\
\end{tabular}
\caption{\label{tab:mass} Physical properties of the sample. Halo mass corresponds to the redshift of the arcs; see text for details. The diameter, $\Delta V$, and $\sigma$ listed for J1038 North refer to the entire J1038 system.}
\end{table*}

\subsection{Gas Fraction}\label{sec:fgas}

The gas mass in star forming galaxies at redshift $z\sim2$ comprises a significant fraction $\sim50$\% of their total baryonic mass \citep{Tacconi10,Daddi10}. Ideally we would like to determine the gas mass from direct observations of atomic H{\sc i} and molecular CO emission. However, in the absence of such data we can estimate total gas mass from the Kennicutt-Schmidt relation. We adopt the best-fit relation of \cite{Kennicutt98}, 
\begin{equation}\label{eq:ks}
\Sigma_{\mathrm{gas}} = (4.0\times10^3 \, \Sigma_{\mathrm{SFR}})^{0.71} \, \mathrm{M_{\odot} pc}^{-2}
\end{equation}
where $\Sigma_{\mathrm{SFR}}$ is in units of \Msunyr kpc$^{-2}$.
The baryonic gas fraction is then defined as
\begin{equation}\label{eq:fgas}
f_{\mathrm{gas}} = \frac{M_{\mathrm{gas}}}{M_{\mathrm{gas}} + M_*}.
\end{equation}
We calculate the gas mass and gas fraction of all three non-merging galaxies using \Ha-derived star formation rates and diameters (Table~\ref{tab:mass}) to determine surface densities. The results are given in Table~\ref{tab:mass}. The inferred gas fractions of $0.4-0.8$ are in good agreement with direct measurements at high redshift \citep{Tacconi10,Daddi10}. Dynamical masses derived in \S\ref{sec:kinematics} also suggest $f_{gas} \approx \frac{M_{dyn}-M_*}{M_{dyn}} = 0.4-0.8$ for the arcs, in excellent agreement.

In cases where high resolution {\em HST} infrared imaging is available, we calculate the spatially resolved gas fraction of the lensed galaxies. For each source-plane pixel we determine $\Sigma_{SFR}$ from \Ha\, intensity and the extinction factors in Table~\ref{tab:mass}, and calculate $\Sigma_{\mathrm{gas}}$ from Equation~\ref{eq:ks}.
The stellar mass density is calculated for the same pixels from the {\em HST} F160W image, which we reconstruct in the source plane and smooth to match the resolution of the OSIRIS \Ha\, data. We assume a constant stellar mass-to-light ratio in the F160W bandpass to calculate $\Sigma_{M_*}$. We extract the average gas and stellar mass densities in radial bins along a psuedo-slit oriented along the kinematic major axis. (The same slits are later used to measure kinematics and metallicity gradients.) We examined the gas fractions as a function of radius and found it to be nearly constant ($<15$\% variation).

\subsection{Emission Line Fitting}\label{sec:fits}

The analysis presented in later sections is based on the resolved properties of emission lines in the galaxy source plane. We determine line flux, velocity, and velocity dispersion from Gaussian fits to the emission lines of interest. The emission line fits for J1206 are described in detail in \cite{Jones10b}. We follow identical methods for the remaining sources presented here, except that emission lines are fit in the source plane rather than the image plane. Here we describe the complete process.

The OSIRIS data for each object is aligned with high resolution {\sl HST} images used to define the lens model transformations from image to source plane (\S\ref{sec:lens}). Each data cube is reconstructed in the source plane and smoothed with a Gaussian kernel to increase signal-to-noise of the fainter emission lines and in regions of low surface brightness. The smoothing kernels have FWHM $\simeq 300$ pc for J0744 and J1038, and 600 pc for J1148. 
We then fit a Gaussian profile to the \Ha\, line at each spatial pixel using a weighted $\chi^2$ to account for increased noise at the wavelengths of bright sky lines. Weights are determined from the noise measured in a blank sky region of each data cube. Line flux and velocity are derived from the centroid and area of the best fit profile in all cases with signal-to-noise $\geq 5$. Intrinsic velocity dispersion is calculated by subtracting the instrumental resolution $\sigma_{inst} \simeq 50$ \kms (measured from bright OH sky lines) in quadrature from the best-fit line width.

We fit all other emission lines of interest using the velocity and dispersion derived from \Ha\, as constraints. \Ha\, is used because it has the highest signal-to-noise in all cases. The centroid and width of each line is fixed at the best-fit values measured for \Ha, and a weighted fit is used to determine the normalization. Each emission line is also fit separately without fixing the centroid and width, and no significant differences are found. We therefore use the emission line fluxes derived with \Ha\, line profiles since these have lower formal uncertainties. Additionally, when considering the ratio of various emission lines, we can be confident that ratios are not biased by different kinematic structure in different lines.

We estimate the physical resolution from observations of tip/tilt reference stars taken at the same time as the emission line data. Tip/tilt stars are reconstructed and smoothed in the same manner and the resulting FWHM is measured along the major and minor axes of reconstructed star images. These values are reported in Table~\ref{tab:obs}. Although smoothing degrades the resolution, the FWHM is $\lsim 1$ kpc for all sources.

\section{Kinematics}\label{sec:kinematics}

\begin{figure*}
\centerline{\includegraphics[width=0.96\textwidth]{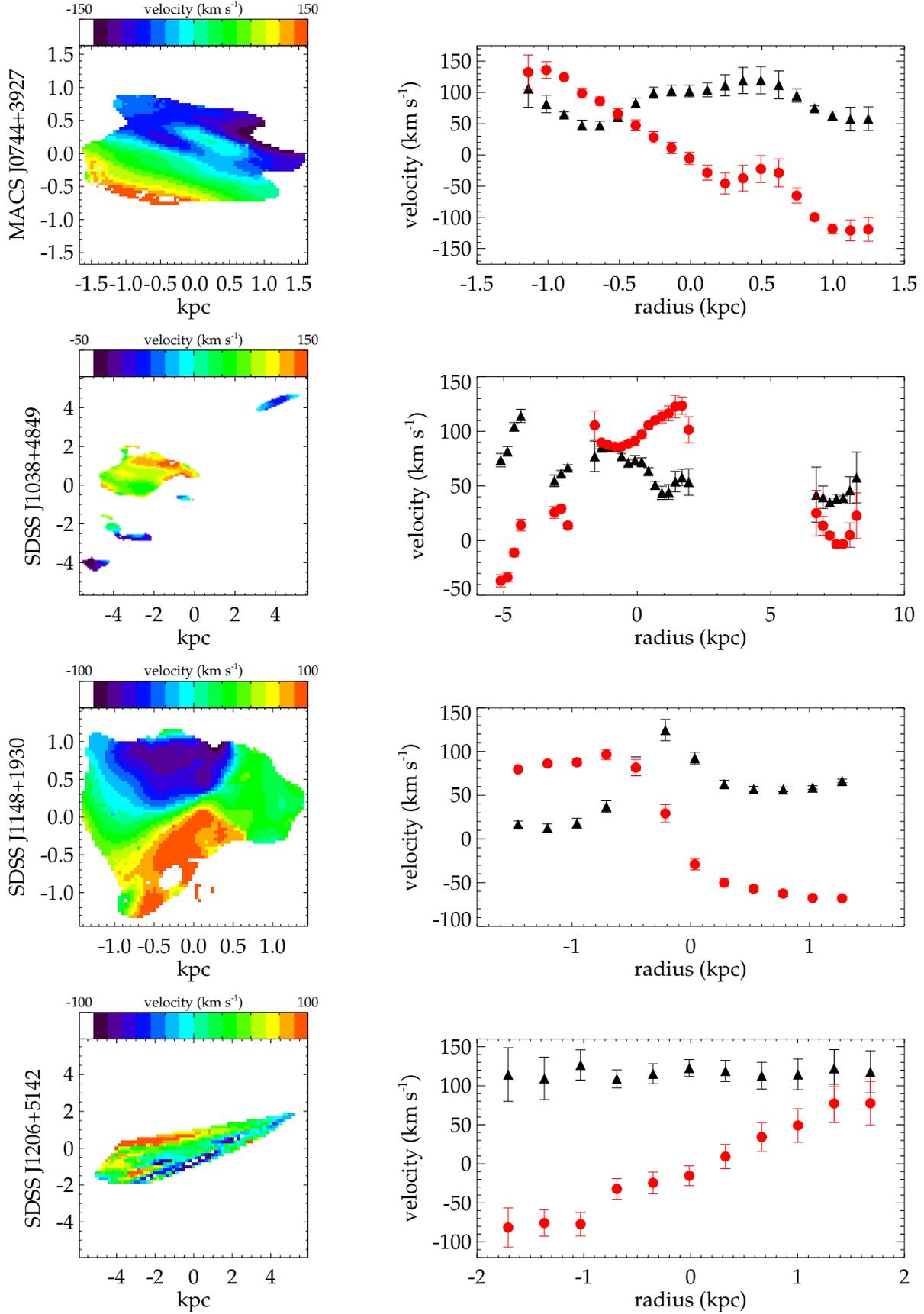}}
\caption{\label{fig:kinematics} 
Source plane kinematics of the lensed galaxies. Left: two-dimensional velocity field. Right: one-dimensional velocity (red circles) and dispersion (black triangles) of each source, extracted along the kinematic major axis.
}
\end{figure*}

In order to understand the variation in metallicity gradients we see in our sample, it is useful to characterize each source in terms of its kinematic properties.
The degree of ordered rotation has been determined from fits to the \Ha\, emission line (\S\ref{sec:fits}). The kinematic properties of J0744 and J1206 are already described in \cite{Jones10a} and \cite{Jones10b} respectively. Here we follow a similar analysis for the remaining sources. As in our earlier work, we extract one-dimensional velocity and dispersion profiles in a pseudo-slit oriented along the kinematic major axis (i.e. the direction of highest velocity shear). These profiles are shown in Figure~\ref{fig:kinematics} along with the two-dimensional velocity maps of each source. Three galaxies (J0744, J1148, and J1206) show ordered rotational motion in both the 2-D maps and 1-D profiles with high local velocity dispersion $\gsim 20$ \kms in all cases. The fourth, J1038, is comprised of multiple spatially and kinematically distinct regions indicating that this source is undergoing a merger. The diameter and peak-to-peak velocity shear $\Delta V$ of each source (measured along the pseudo-slit) are given in Table~\ref{tab:mass}. For the rotating sources, $\Delta V$ is related to the maximum circular velocity $V_{max}$ and inclination angle $i$ as 
\begin{equation}
\label{eq:dv}
\Delta V = 2 V_{max} \sin{i}.
\end{equation}
Additionally we give the mean local velocity dispersion $\sigma$, defined as the unweighted mean of individual pixels with error bars reflecting the $1\sigma$ scatter. We note that the adopted definition of $\sigma$ varies throughout the relevant literature.

We now briefly compare the kinematics with other samples at high redshift reported in the literature. The largest such sample is the SINS survey; observations of 80 SINS galaxies at $z\sim2$ are described in \cite{Forster09}. The median velocity shear of their sample is $\Delta V = 134$ \kms with an interquartile range of $90-220$ \kms. Approximately one third of the SINS sources are clasified as mergers while the rest have varying degrees of rotation and random motion. \cite{Forster09} tabulate the ratio of velocity shear to velocity dispersion, $\Delta V / (2 \sigma)$, which gives an indication of the degree to which each galaxy is dynamically supported by rotation vs. random motions. The SINS sample has a median $\Delta V / (2 \sigma) = 0.56$ and interquartile range $0.37-0.75$. Other published samples of $z\simeq2-3$ galaxies have median $\Delta V / (2 \sigma)$ values of 0.4 \citep{Law09} and 0.9 \citep{Jones10a}. 
In comparison, velocity shear observed in the lensed galaxies presented here is somewhat higher than in other samples at similar redshift. The values in Table~\ref{tab:mass} give a range $\Delta V / (2 \sigma) = 0.76-1.4$ for our lensed galaxies, and we note that our values of $\sigma$ are systematically higher than for the definition used by \cite{Forster09}. The lensed galaxies therefore have a higher degree of rotation than typical SINS sources. According to the criteria of \cite{Forster09}, all three non-merging sources in our sample are ``rotation-dominated'' (defined as having $\Delta V / (2 \sigma) \geq 0.4$).

A common hypothesis advanced by recent studies is that the evolution of star-forming galaxies at high redshift is driven in large part by global gravitational instability (e.g. \citealt{Jones10a, Genzel08, Dekel09}). Briefly, if the velocity dispersion and rotational velocity are too small, the system is unstable and gravitational perturbations will grow exponentially into giant ``clumps''. This is quantified by the Toomre parameter $Q$, where values $Q \lsim 1$ indicate that a galaxy is unstable and will fragment into giant clumps. We calculate $Q/\sin{i}$ for the rotating galaxies in our sample using the information in Table~\ref{tab:mass} and following the same method as \cite{Jones10a}. We list the values in Table~\ref{tab:mass} as well as the dynamical mass estimated as $M_{dyn} = 5 R \sigma^2 / G$. The results indicate that all three galaxies are gravitationally unstable unless the inclinations are very high ($> 60$ degrees), although the uncertainty is approximately a factor of 2 in $Q$ (see \citealt{Jones10a}). This can explain the ``clumpy'' morphologies apparent in Figure~\ref{fig:morphology}.

In summary, three galaxies in our lensed sample are rotating and one is undergoing a merger. This merger fraction is consistent with larger samples at similar redshift. The kinematics of galaxies in our sample are more rotation-dominated than others studied at similar redshift, and all rotating galaxies in our sample are likely to be gravitationally unstable. The ratio of shear to velocity dispersion, $V/\sigma$, is $\sim2\times$ higher than for typical galaxies in the extensive SINS survey.

\section{Gas Phase Metallicity}\label{sec:metallicity}

The multiple emission lines detected by OSIRIS enable us to determine spatially resolved gas-phase metallicities in each galaxy. Robust metallicity measurements rely on knowledge of the temperature and ionization state of a gas. However, this requires accurate measurement of diagnostic emission lines which are usually too faint to be detected at high redshift with current instruments. Instead we use ratios of strong emission lines which can be detected in modest integration times and which correlate with metallicity through locally-calibrated relations. These include ratios such as $\frac{\mbox{\Oiii + \Oii}}{\mbox{\Hb}}$ (the R23 index) and \Nii$/$\Ha\, (the N2 index) for which calibrations have been determined using direct electron temperature-based measurements or photoionization models (e.g. \citealt{Pagel79, Kewley02, Pettini04, Maiolino08}). As in \cite{Jones10b} we primarily use the N2 index to determine gas-phase oxygen abundance via
\begin{equation}\label{eq:n2}
12 + \log{O/H} = 8.90 + 0.57 \times \log{[N II]/H\alpha}
\end{equation}
\citep{Pettini04}, with a dispersion of 0.18 dex.
The main advantage of N2 is that the two requisite lines are very close in wavelength such that systematic uncertainties from reddening and instrumental effects are negligible.

While the N2 index provides a practical means of estimating metallicity in high-redshift galaxies, it can fail in cases where ({\em i}) active galactic nuclei (AGN) or shock excitation contribute significantly to the emission line flux, ({\em ii}) secondary production of nitrogen leads to variations in the N/O ratio, or ({\em iii}) \Nii\, cooling saturates (at $12+\log{O/H} \gsim 9.0$). This last case is not a concern for the galaxies in our sample as they have metallicities significantly below this value. However AGN, shocks, and variations in the N/O ratio are a potential problem and we address this issue with additional observations of \Oiii\, and \Hb\, emission lines.

\begin{figure}
\centerline{\includegraphics[width=0.5\textwidth]{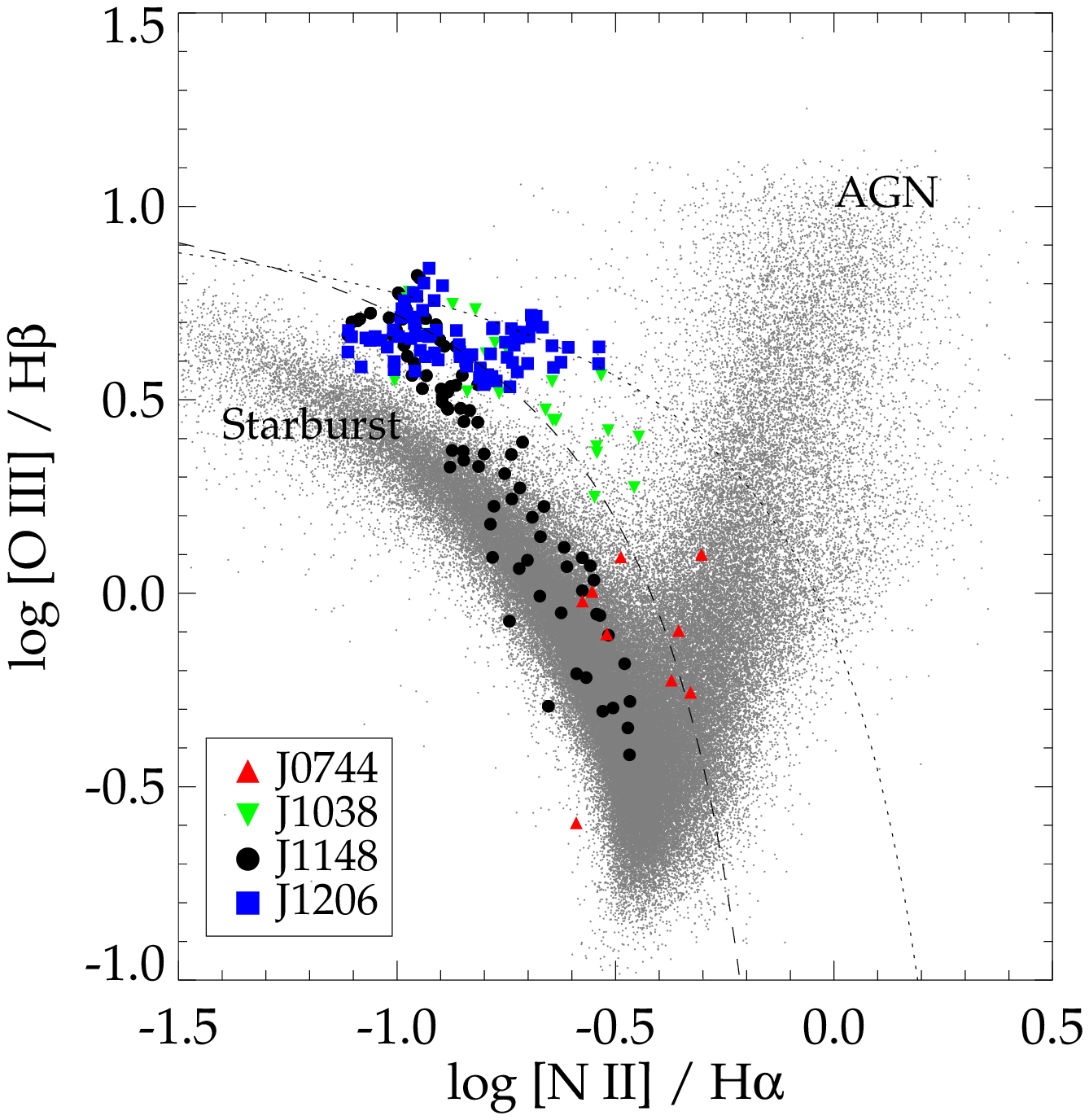}}
\caption{\label{fig:bpt} 
Diagnostic diagram of \Nii/\Ha\, and \Oiii/\Hb. Each large point represents a single spatial pixel of OSIRIS, and the small grey points are $\sim$122,000 galaxies from the Sloan Digital Sky Survey (SDSS) with signal to noise $\geq5$ in all relevant emission lines. SDSS galaxies show a locus of star forming galaxies, and a separate branch of AGN extending to the upper right. The dotted line shows the theoretical maximum \Oiii/\Hb\, from star forming regions \citep{Kewley01}, and the dashed line is an empirical division between star-forming galaxies and AGN from \cite{Kauffmann03}. All regions of the lensed galaxies are consistent with pure star formation, although \Oiii/\Hb\, is typically above the locus of SDSS star forming galaxies at $z\simeq0$. This is commonly observed at high redshift and is attributed to a high ionization parameter (e.g. \citealt{Hainline09,Erb10}).
}
\end{figure}

AGN and shocks can be distinguished from star-forming regions on the basis of emission line flux ratios. One of the most widely used diagnosics is the ratio of \Nii/\Ha\, compared to \Oiii/\Hb, as described by \cite{Baldwin81} (the ``BPT diagram''). We show the BPT diagram for individual pixels in each lensed galaxy in Figure~\ref{fig:bpt}. Only regions where all requisite lines are detected ($>3\sigma$) are shown on the diagram. We note that some individual pixels are correlated since the point spread function is larger than the pixel size.
In the case of J1206 we estimate \Hb\, flux from the global ratio \Ha/\Hb~$= 4.07$ \citep{Hainline09}; these \Oiii/\Hb\, values are therefore accurate on average but may deviate significantly within individual pixels. The line ratios of $\sim$122,000 galaxies from the Sloan Digital Sky Survey (SDSS) are also plotted and clearly show a locus of star forming galaxies with a separate branch of AGN. The lensed galaxies generally follow the locus of local star forming galaxies but are offset to higher \Oiii/\Hb\, ratios; this has been commonly observed in high redshift galaxies and is generally attributed to high ionization parameters rather than AGN (see discussion in \citealt{Erb10} and \citealt{Hainline09}). Some regions of the lensed galaxies have \Oiii/\Hb\, formally above the theoretical limit from star formation \citep{Kewley01}, but consistent within the uncertainties. We conclude that emission lines in the lensed galaxies originate predominantly from star forming H{\sc ii} regions with little contribution from AGN or shocks.

\begin{figure*}
\centerline{\includegraphics[width=0.36\textwidth]{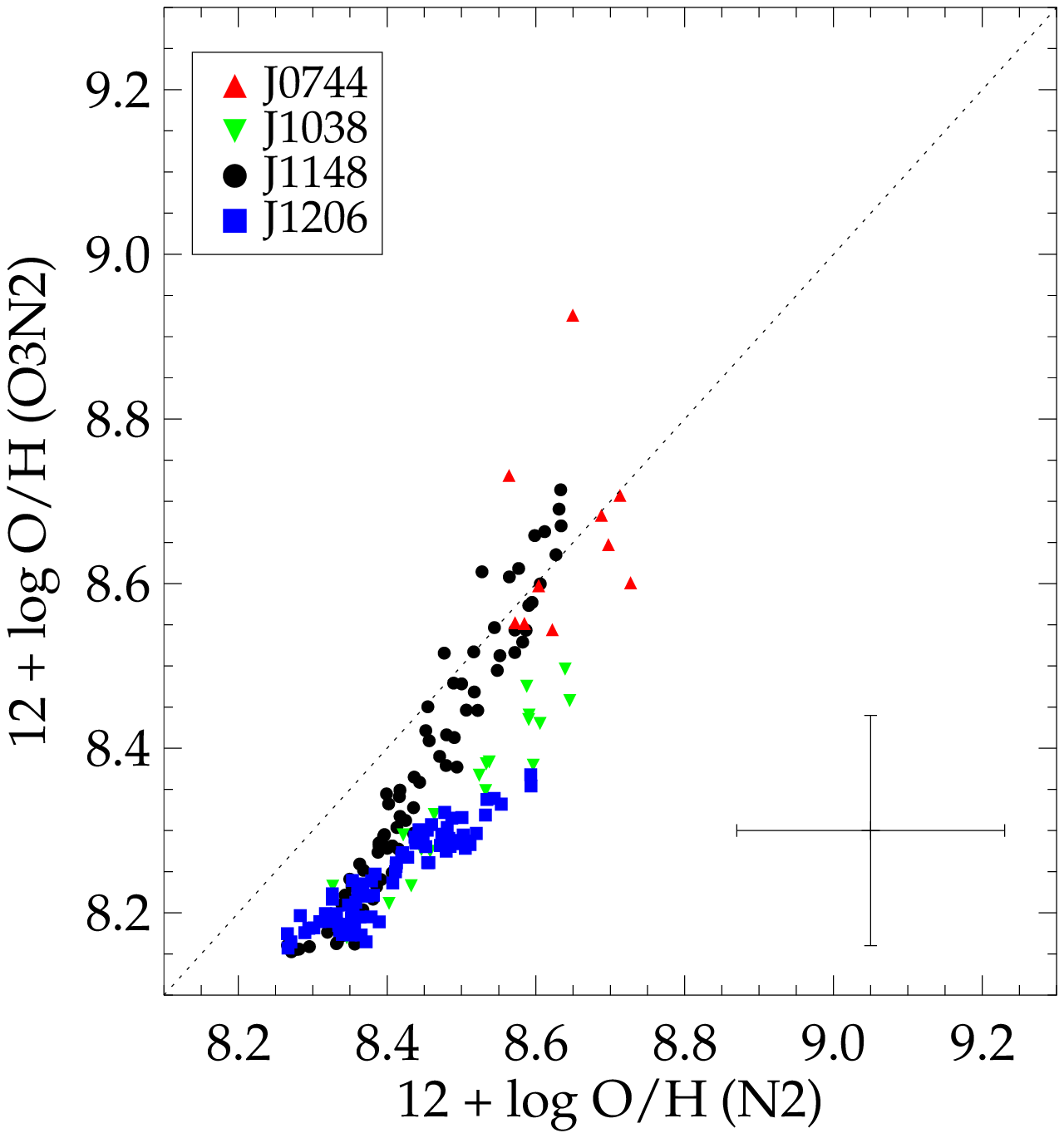}
\hspace{-.04\textwidth}
\includegraphics[width=0.36\textwidth]{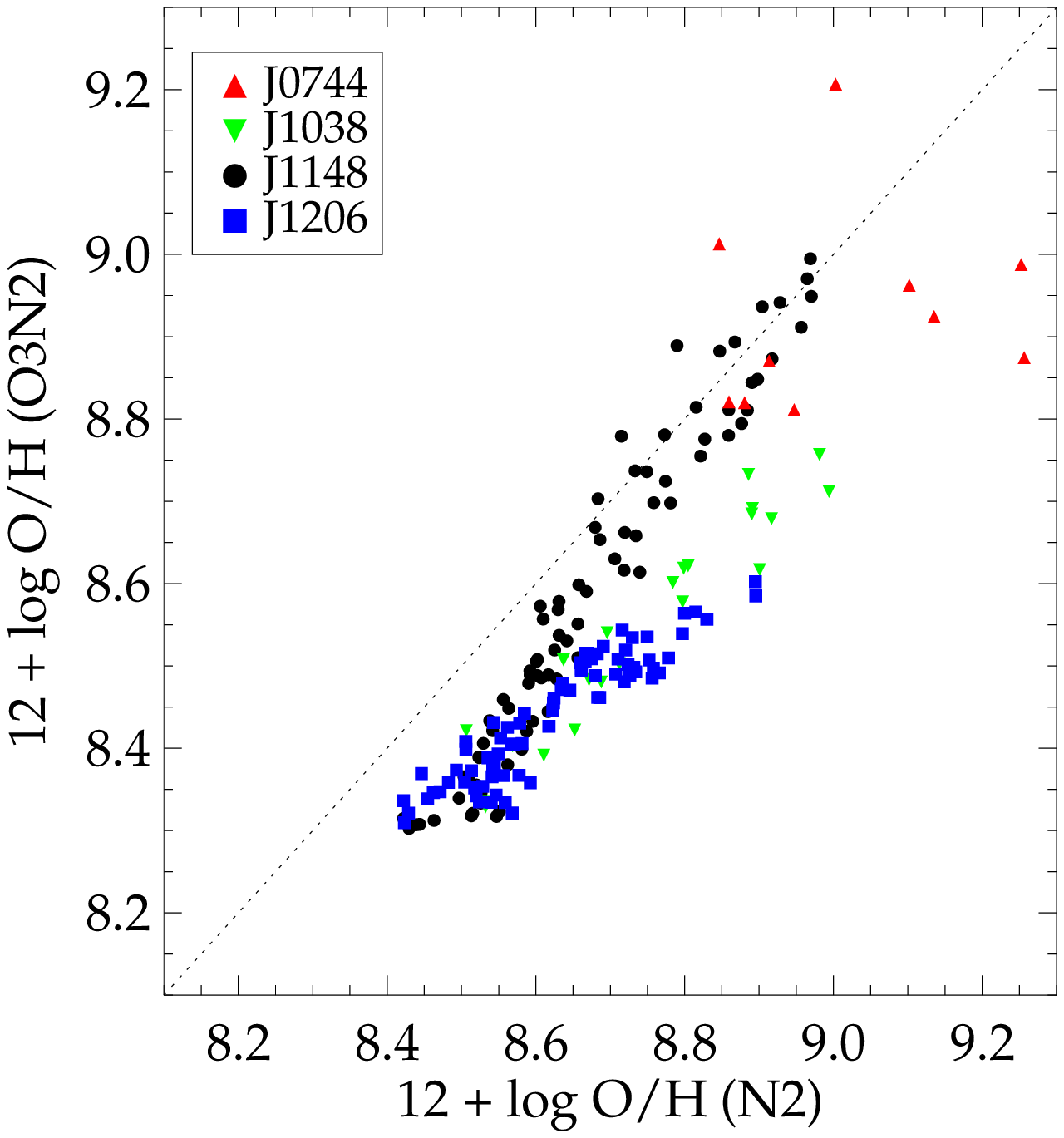}
\hspace{-.04\textwidth}
\includegraphics[width=0.36\textwidth]{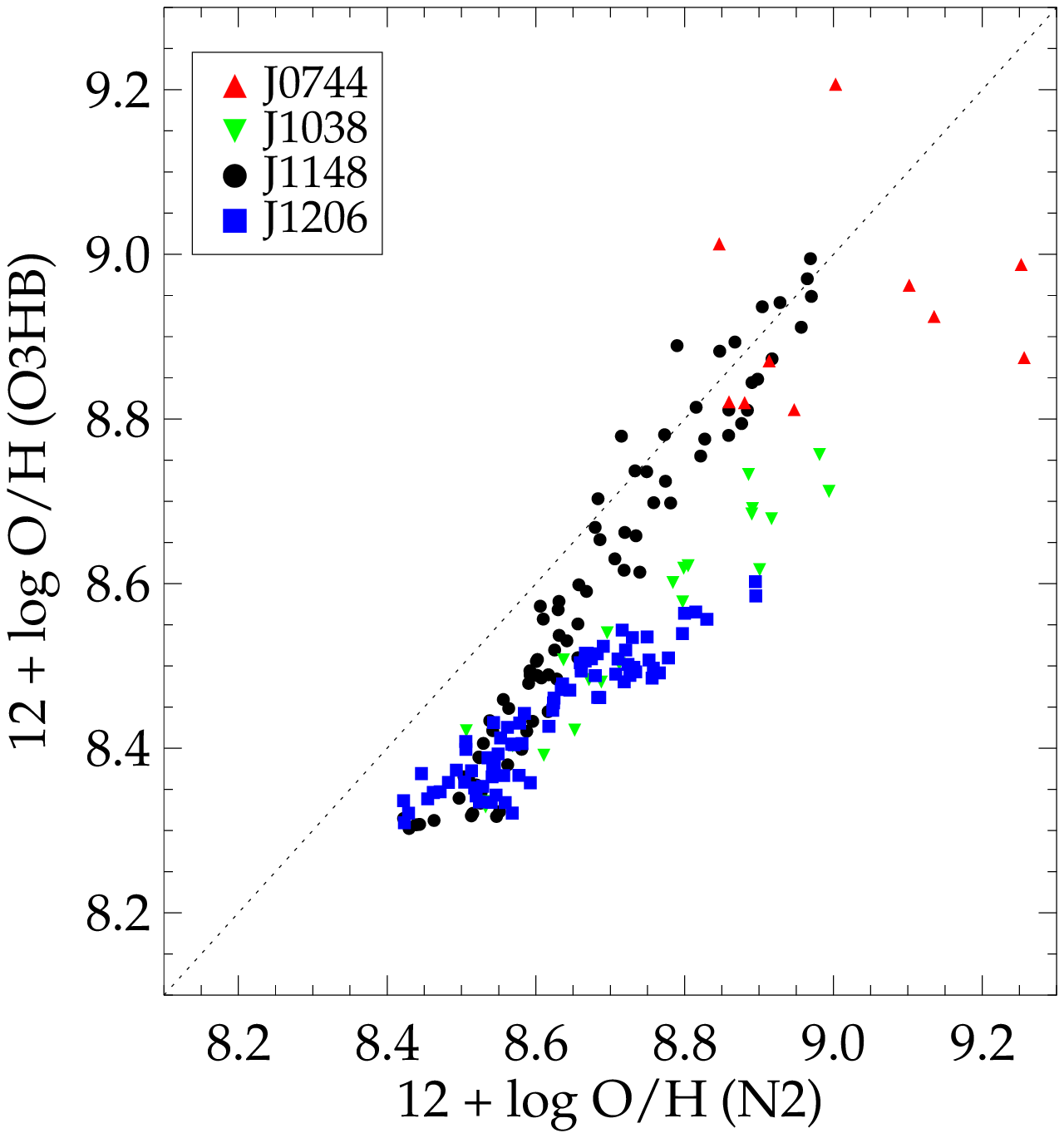}}
\caption{\label{fig:metallicity_compare} 
Comparison of the different strong-line metallicity diagnostics N2 (\Nii/\Ha), O3N2 (\Oiii/\Nii), and O3HB (\Oiii/\Hb). Metallicities are derived from the calibrations of \cite{Pettini04} (left panel) and \cite{Maiolino08} (center and right panels). Each point represents an individual OSIRIS pixel; data are identical to those in Figure~\ref{fig:bpt}. The $1\sigma$ scatter measured by \cite{Pettini04} is shown in the lower right of the left hand panel. Metallicities derived from different methods are generally in agreement, with \Oiii-based metallicities typically $0.15$ dex lower than those derived from N2. Metallicity derived using the \cite{Maiolino08} calibrations are systematically higher by $\sim0.2$ dex compared to \cite{Pettini04}.
}
\end{figure*}

To test whether metallicities derived from the N2 index suffer from systematic errors due to variable N/O abundance or any other effect, we calculate metallicities using various other methods. Using the same data as for the BPT diagram (Figure~\ref{fig:bpt}), we compute 12+log(O/H) using the calibrations of \cite{Pettini04} and \cite{Maiolino08} and show the results in Figure~\ref{fig:metallicity_compare}. All metallicities are consistent given the intrinsic scatter in each calibration, although \Oiii-based metallicities are typically lower by $\sim 0.15$ dex than those inferred from \Nii\, alone. Furthermore the \cite{Maiolino08} results are systematically higher than for \cite{Pettini04} by $\sim 0.2$ dex. Overall we conclude that systematic uncertainties in metallicity calculated from the N2 index are limited to $\lsim 0.2$ dex depending on the calibration used (e.g. N2 vs. O3HB), with an additional $\sim 0.2$ dex uncertainty in the zero point (e.g. from \citealt{Pettini04} vs. \citealt{Maiolino08}). In the following section we show how different calibrations affect the derived metallicity gradients.

\begin{figure*}
\centerline{\includegraphics[width=0.95\textwidth]{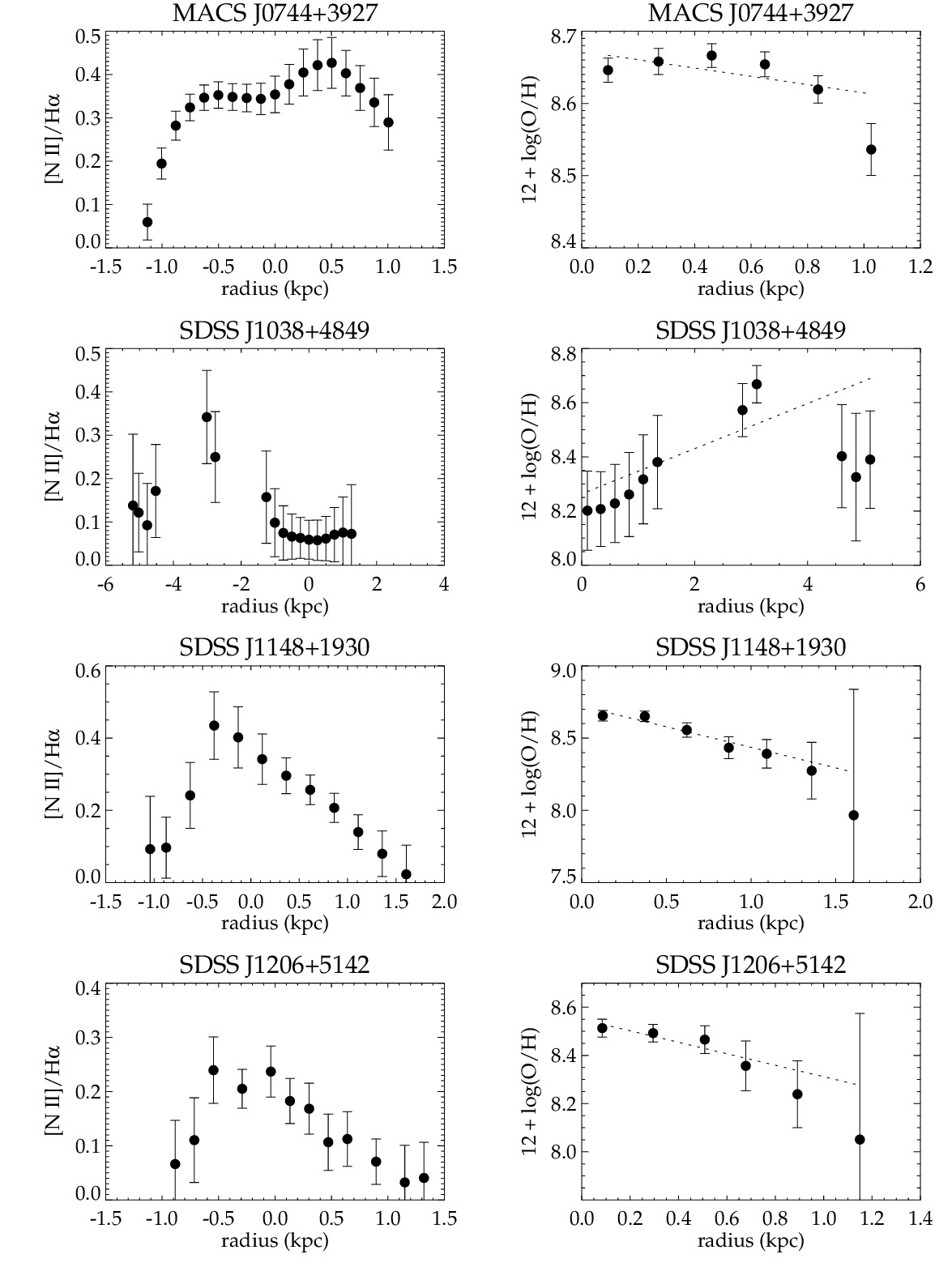}}
\caption{\label{fig:gradients} 
Metallicity gradients of the lensed galaxies. Left: \Nii$/$\Ha\, ratio as a function of radius. Right: Gas-phase metallicity as a function of radius derived from the N2 index calibration of \cite{Pettini04}. Dashed lines show the best fit linear metallicity gradient. \Nii$/$\Ha\, ratios are extracted along the kinematic major axis using the same slit as for the right panels of Figure~\ref{fig:kinematics}.
}
\end{figure*}

\subsection{Metallicity Gradients}

We are now in a position to determine gradients in the inferred metallicity along the kinematic major axis of each galaxy. To do so we extract resolved measurements of \Nii\, and \Ha\, flux along a pseudo-slit in the same manner as for constructing the rotation curves shown in Figure~\ref{fig:kinematics}. While other orientations can also be used, they must be corrected by the uncertain inclination and so we take the major axis to provide the most precise measurement. In the case of J1038 the inclination angle between merging components is unknown, hence the magnitude of gradients for this source should be considered an upper limit. The center of each galaxy is defined kinematically as the position along the pseudo-slit at which the velocity is equal to the systemic velocity (i.e. velocity $=0$ in Figure~\ref{fig:kinematics}); for the merging system J1038 we use the systemic velocity of the UV-bright northern region (corresponding to velocity $\simeq 100$ \kms\, in Figure~\ref{fig:kinematics}).
We bin individual pixels by their galactocentric radius and compute the total \Ha\, and \Nii\, flux using all pixels with signal-to-noise $\geq 10$ in \Ha. Metallicity is derived from the resulting \Nii$/$\Ha\, ratio via Equation~\ref{eq:n2} and shown as a function of radius in Figure~\ref{fig:gradients}.

Each galaxy shows significant variations in metallicity as a function of radius (Figure~\ref{fig:gradients}). The three rotating galaxies (J0744, J1148, and J1206) all show significant gradients with decreasing metallicity at larger radii, as observed in all local disk galaxies. The merging system J1038 has significantly lower nuclear metallicity than the rotating galaxies, and exhibits a more complex metallicity distribution. Metallicity is measured for three spatially distinct regions of J1038. The peak of rest-UV and \Ha\, emission (at $R=0$ in Figure~\ref{fig:gradients}) has the lowest metallicity at $12+\log{O/H}=8.2$, while the peak of IRAC 3.6$\mu$m flux (at $R=5$ kpc) has marginally higher $12+\log{O/H}=8.4$. The third component, located between the first two (at $R=3$ kpc) has the highest measured metallicity.

We now quantify the radial metallicity gradient of each source in order to compare the results with other samples. Gradients in local galaxies are commonly expressed as a linear relation in units of dex$/$kpc. We compute the best linear fit to $12+\log{O/H}$ as a function of radius using a weighted least-squares method. The best fit relations are shown as dashed lines in Figure~\ref{fig:gradients} (right panels) and provide a reasonable fit to the data with reduced $\chi^2$ values of $0.3-2.3$, although we note that the linear fits overpredict metallicity at the largest radii in all sources. The best-fit central metallicity and gradient of each source is listed in Table~\ref{tab:gradients2}.

\begin{figure*}
\centerline{\includegraphics[width=\textwidth]{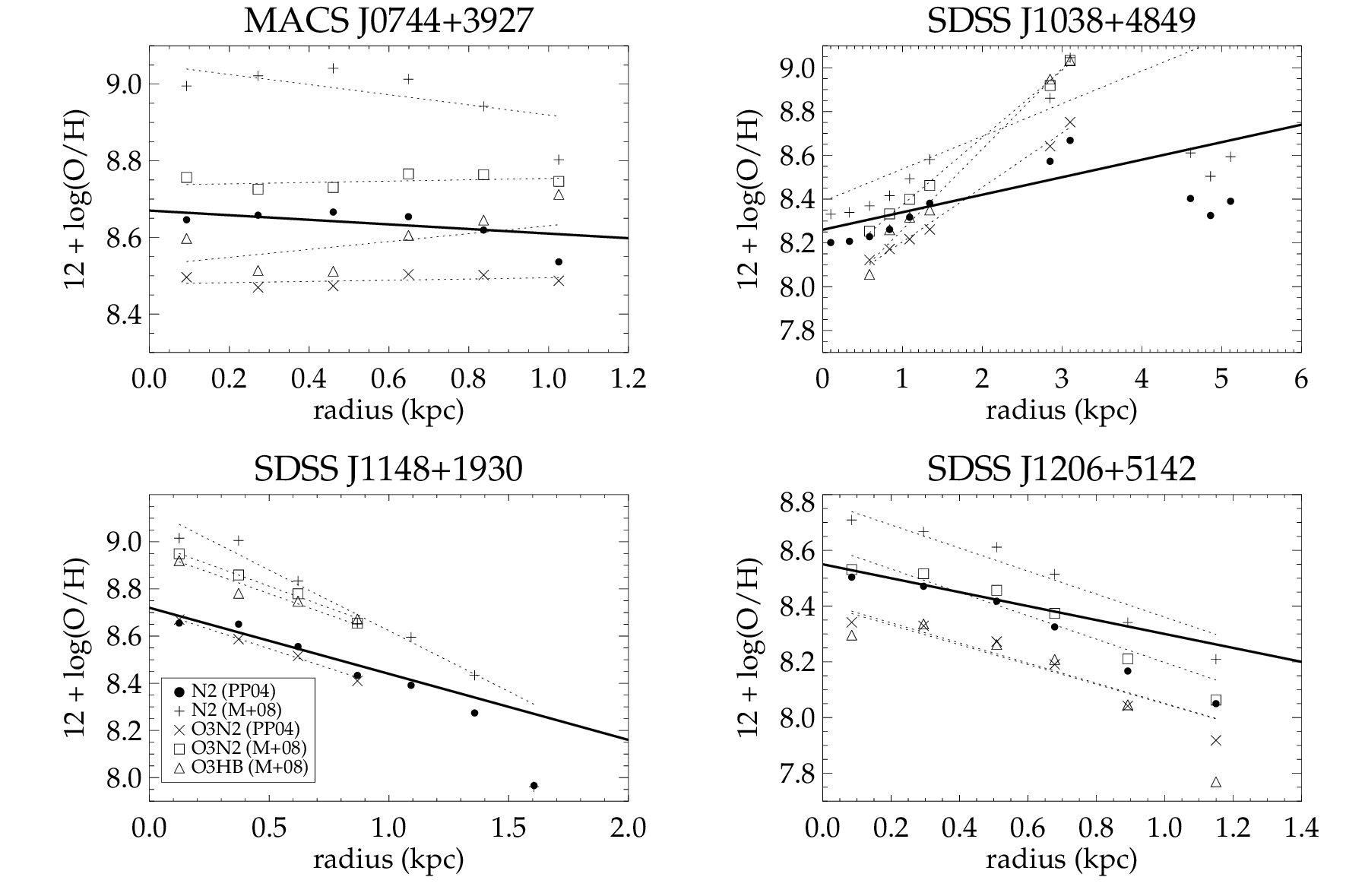}}
\caption{\label{fig:gradients_compare} 
Metallicity gradients for each galaxy in our sample based on the use of various strong-line diagnostics. N2, O3N2, and O3HB are as presented in Figure~\ref{fig:metallicity_compare}, and we use the calibrations of \cite{Pettini04} (PP04) and \cite{Maiolino08} (M+08). Solid lines show the gradient from Figure~\ref{fig:gradients} and dotted lines are the best fits to the other diagnostics. For J0744 and J1206, we estimate \Hb\, from \Ha\, with the global \Ha/\Hb\, ratio. In nearly all cases the gradients are approximately equal to or stronger than the adopted value (N2, PP04). The exceptions are the \Oiii-based results for J0744 which suggest a flat or inverted gradient, although all slopes are within $2\sigma$ of the adopted value.
}
\end{figure*}

To examine whether these results might be affected by systematic calibration errors, we compute the metallicity gradient in the exact same manner using different strong-line metallicity diagnostics. In Figure~\ref{fig:gradients_compare} we show the radial metallicity profiles determined from all available calibrations described by \cite{Pettini04} and \cite{Maiolino08} as well as their best-fit linear gradients. We note that J1206 is assumed to have a constant \Ha/\Hb\, ratio, that \Oiii\, and \Hb\, in J1038 were only observed in the region corresponding to $R<4$ kpc, and that \Oiii\, and \Hb\, are only detected with sufficient signal-to-noise at $R<1$ kpc in J1148. We further assume a constant \Ha/\Hb\, ratio for J0744 since \Hb\, is detected in only a few pixels. Systematic offsets between the various calibrations are apparent in Figure~\ref{fig:gradients_compare}, but in general the different methods agree given their intrinsic scatter. We give the best-fit gradients in Table~\ref{tab:gradients2}. In most cases the resulting gradients are {\em stronger} than that for our adopted N2 calibration, and in all cases the uncertainty is larger. Importantly, \Oiii-based calibrations (including O3HB which is independent of N2) confirm that line ratio gradients are due to metallicity gradients rather than N/O abundance or other effects. The exception is J0744 for which \Oiii\, and \Hb\, ratios suggest a positive metallicity gradient. Overall, the different metallicity derivations are in reasonable agreement, albeit with large uncertainties due to both low signal-to-noise and intrinsic scatter in the calibrations. In the following sections we will consider the N2-based gradients of each galaxy to be the most accurate.

\begin{table*}
\begin{tabular}{lcccccc}
Name        &  Central $12+\log{O/H}$    &   N2 PP04           &   N2 M+08        &  O3N2 PP04       & O3N2 M+08        &  O3HB M+08       \\
            &  (from N2, PP04)        &  (dex/kpc)          &  (dex/kpc)       &  (dex/kpc)       &  (dex/kpc)       &  (dex/kpc)       \\
\hline
MACS J0744  &  $8.67 \pm 0.18$  &  $-0.06 \pm 0.04$   & $-0.13 \pm 0.05$ & $0.02 \pm 0.04$  & $0.02 \pm 0.04$  & $0.10 \pm 0.07$  \\
SDSS J1038  &  $8.26 \pm 0.19$  &  $0.08 \pm 0.03$    & $0.15 \pm 0.07$  & $0.25 \pm 0.07$  & $0.31 \pm 0.08$  & $0.37 \pm 0.09$  \\
SDSS J1148  &  $8.72 \pm 0.18$  &  $-0.28 \pm 0.05$   & $-0.51 \pm 0.11$ & $-0.33 \pm 0.12$ & $-0.37 \pm 0.11$ & $-0.37 \pm 0.19$ \\
SDSS J1206  &  $8.55 \pm 0.18$  &  $-0.25 \pm 0.06$   & $-0.45 \pm 0.08$ & $-0.40 \pm 0.08$ & $-0.46 \pm 0.08$ & $-0.40 \pm 0.11$ \\
\end{tabular}
\caption{\label{tab:gradients2} 
Metallicity gradients derived from different strong-line diagnostics shown in Figure~\ref{fig:gradients_compare}.
}
\end{table*}

\subsection{Evolution with Redshift}\label{sec:z_evo}

\begin{figure*}
\centerline{\includegraphics[width=\textwidth]{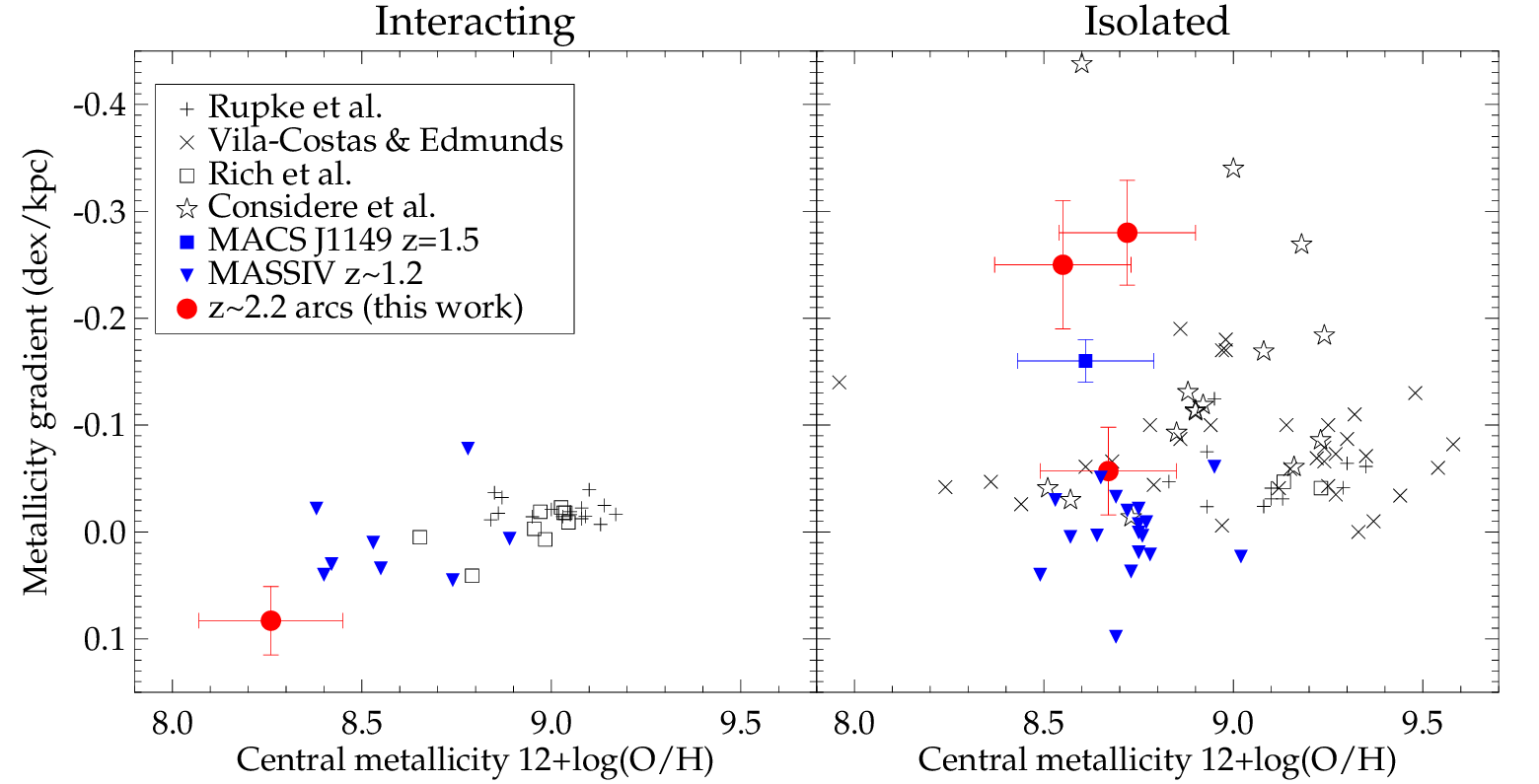}}
\caption{\label{fig:gradients2} 
Metallicity gradients of the lensed galaxies compared with various other published samples (\citealt{Rupke10,Vila-Costas92,Rich12,Considere00}; MACS J1149: \citealt{Yuan11}; MASSIV: \citealt{Queyrel12}). The plot is divided into interacting (or merging) systems and isolated galaxies. All data points at high redshift are shown as filled color symbols. Interacting galaxies have generally lower metallicities and relatively flat or inverted gradients ($\lsim-0.05$ dex kpc$^{-1}$) compared to isolated galaxies.
}
\end{figure*}

The metallicity gradients we have determined for our lensed sample can now be compared to those for galaxies at different redshifts. Figure~\ref{fig:gradients2} shows a comparison with other measurements reported in the literature. It is instructive to separate the sample according to whether the galaxies are isolated or interacting. Figure~\ref{fig:gradients2} 
demonstrates that interacting galaxies tend to have flatter gradients and lower central metallicity. Two of the lensed $z\sim2$ sources, J1148 and J1206, have gradients $<-0.2$ dex/kpc which are significantly steeper than those found in local massive disk galaxies. The only local galaxies in Figure~\ref{fig:gradients2} with similar or steeper gradients are barred spiral starbursts from \cite{Considere00} and dwarf galaxies from \cite{Vila-Costas92}. J0744 has a gradient typical of local disk galaxies. The merging galaxy J1038 has an ``inverted'' gradient ($>0$ dex/kpc) - a phenomenon which has also been observed in local mergers \citep{Rich12} and high-redshift sources \citep{Cresci10, Queyrel12}. 

\begin{figure*}
\centerline{\includegraphics[width=\textwidth]{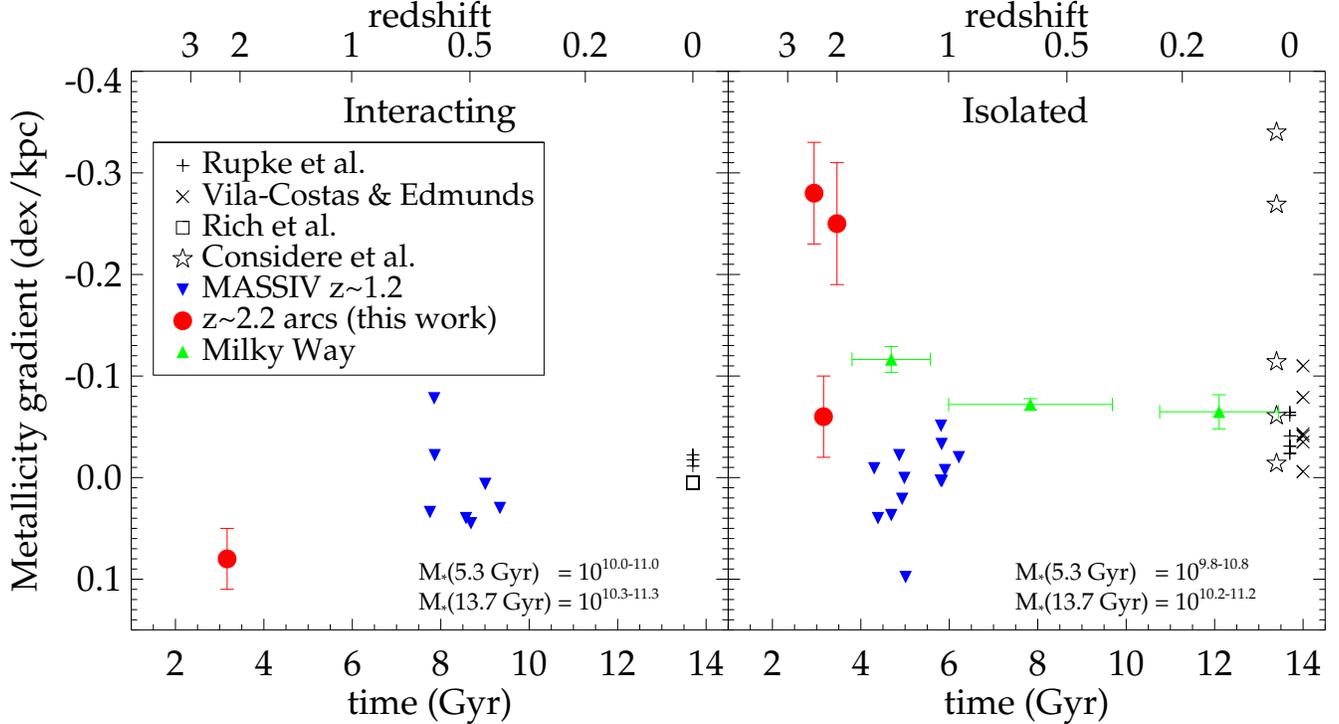}}
\caption{\label{fig:gradients3}\label{fig:gradients4}
Metallicity gradients of the lensed galaxies (red circles) compared to equivalent samples at other redshifts. The x-axis is linear in cosmic time with corresponding redshifts shown on the upper axis.
The galaxies shown here are a subset of those in Figure~\ref{fig:gradients2} and symbols are identical. Comparison samples are selected to occupy the same dark matter halos as the lensed galaxies based on galaxy stellar mass; the corresponding stellar masses at $z=1.2$ and $z=0$ are noted at the bottom right of each panel. Samples at $z=0$ are plotted with slight offsets in redshift for ease of viewing.
We also show the time evolution of the Milky Way metallicity gradient measured by \cite{Maciel03} in the appropriate panel.
}
\end{figure*}

Although the overall sample size is still limited, it appears that the metallicity gradients vary systematically with redshift and here we seek to quantify this evolution.
However, before doing so, it is important to consider how best to compare galaxies at different redshifts in a consistent manner. One way to proceed is to determine the approximate stellar mass that the lensed galaxies will have at different cosmic epochs based on halo abundance matching techniques. This enables us to select a comparison sample with stellar masses approximately equal to the value inferred for the lensed galaxies at that redshift. The first step in this process is to estimate the halo mass for our lensed galaxies. Several groups have studied the stellar to halo mass ratio and its variation with redshift by combining cosmological simulations with measured stellar mass functions (e.g. \citealt{Yang11, Conroy09, Moster12}). We use the results of \cite{Moster12} to do this and report the values in Table~\ref{tab:mass}. We calculate the corresponding halo mass at different redshifts by integrating the mean halo growth rate determined by \cite{Fakhouri10}, and determine the stellar mass at these redshifts via the \cite{Moster12} formalism. Next we must determine galaxy stellar masses for the comparison samples. For the MASSIV sample, we use the stellar masses derived by \cite{Queyrel12}. For all other comparison samples at $z\simeq0$, we extract broadband $B-V$ color and absolute luminosity ($M_B$ and/or $M_K$) from the NASA Extragalactic Database and calculate the stellar mass using the mass-to-light ratio formulae from \cite{Torres-Flores11}. Finally, we construct two separate comparison samples appropriate for (1) isolated and (2) interacting galaxies. For isolated galaxies we choose a stellar mass range corresponding to $M_* = 10^{9.7\pm0.5}$ at $z=2.2$; the isolated lensed galaxies are all within 0.3 dex of this value. The stellar mass range for interacting galaxies corresponds to that of J1038. We show the metallicity gradients of each comparison sample as a function of linear cosmic time and equivalent redshift in Figure~\ref{fig:gradients3}.

\begin{figure}
\centerline{\includegraphics[width=0.5\textwidth]{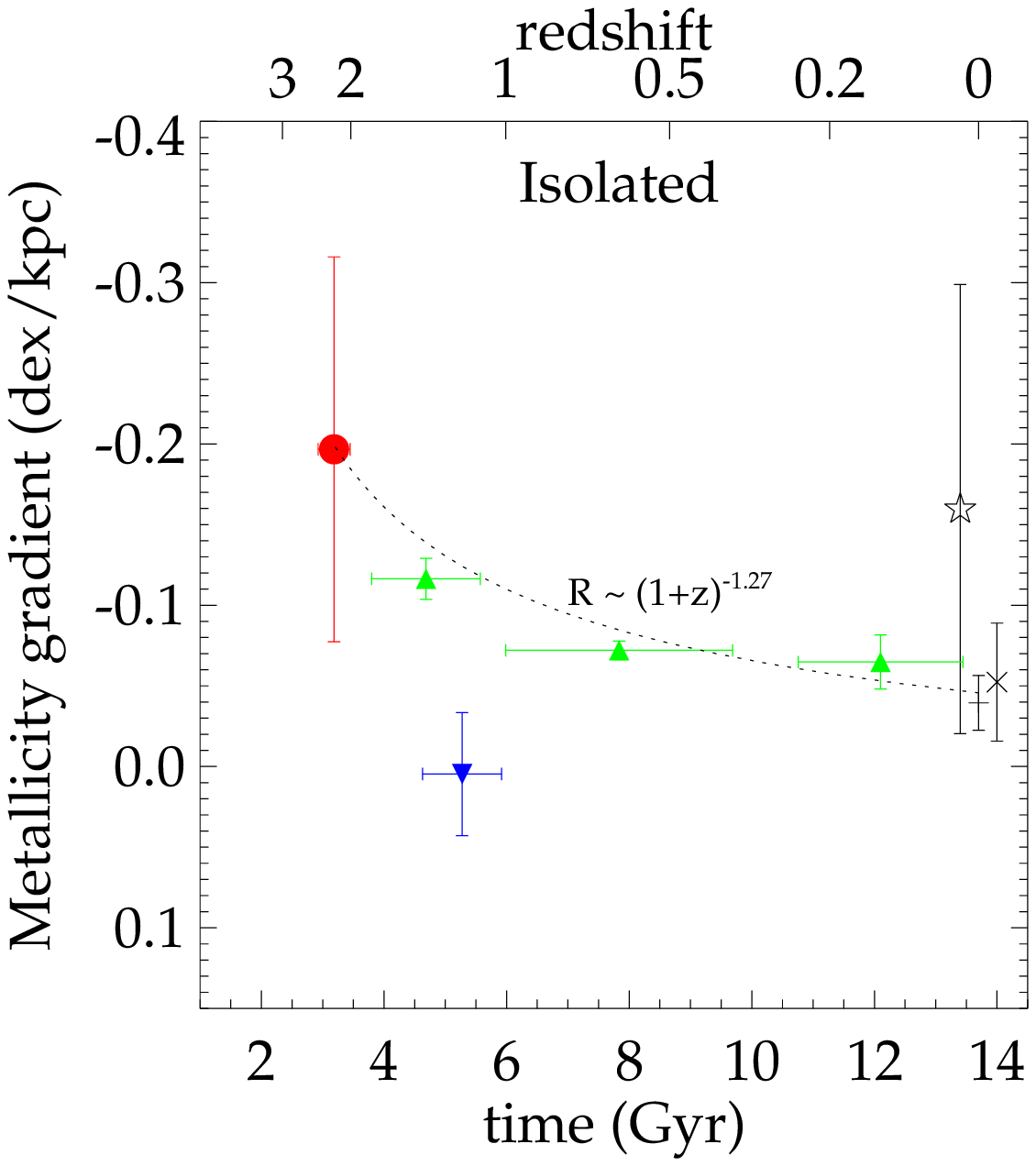}}
\caption{\label{fig:gradients5} 
Equivalent to the right panel of Figure~\ref{fig:gradients3}, showing the mean values of each sample. Error bars denote the $1\sigma$ dispersion within each sample. The dotted line shows how gradients will evolve with time if the range of metallicity remains constant while the characteristic radius grows as $R \propto (1+z)^{-1.27}$ as determined by \cite{vanDokkum10}, scaled to match the lensed galaxies at $z=2.2$.
}
\end{figure}

Figure~\ref{fig:gradients3} shows a clear trend in metallicity gradients with time (or redshift) for our sample of galaxies which occupy similar dark matter halos. The scatter at a given redshift presumably reflects the degree of intrinsic variation within the population. To investigate this with more clarity, we show in Figure~\ref{fig:gradients5} the mean and $1\sigma$ scatter of each isolated galaxy comparison sample. Considering only the $z\geq2$ and $z\sim0$ data in Figure~\ref{fig:gradients5}, we can see that the average metallicity gradient becomes flatter with time although the intrinsic scatter is similar to the average gradient. However, data at $z\sim1.2$ from the MASSIV survey do not support this picture; instead they show a mean gradient $\simeq 0$ with relatively low scatter of $\sim 0.05$ dex. This discrepancy is puzzling and warrants further investigation. Formally the mean metallicity gradient for the samples shown in Figure~\ref{fig:gradients5} is $-0.20 \pm 0.07$ at $z=2.2$, $0.005\pm0.011$ at $z=1.2$, and $-0.077\pm0.005$ at $z=0$. The local data and lensed galaxies suggest that on average, metallicity gradients have declined by a factor of $2.6\pm0.9$ in the past 10 Gyr. Naturally metallicity gradients as steep as $-0.2$ dex/kpc cannot be common in massive galaxies at $z\simeq0$ simply because of their size. In the Milky Way, solar metallicity at $R=8$ kpc would imply $40\times$ solar metallicity in the center with such a gradient. 

We now consider the effect of ``inside-out'' size growth. Supposing that the relative metal enrichment at the center and effective radius $R_e$ of a galaxy remains constant with time, the gradient will scale as $1/R_e$ and will flatten with time as the radius increases. \cite{vanDokkum10} have measured the size of massive galaxies selected to have a fixed number density at different redshifts (equivalently, approximately the same halos), and find $R_e \propto (1+z)^{-1.27}$. We show the appropriate change in metallicity gradient ($\propto (1+z)^{1.27}$) in Figure~\ref{fig:gradients5}, and find that it is within $1\sigma$ of the Milky Way and $z=0$ data. However, the galaxies used to determine this size growth are significantly more massive (by 0.7 dex at $z=0$) than the metallicity gradient sample and likely grow at a different rate. Adopting the same functional form $R_e \propto (1+z)^{-\alpha}$, the lensed galaxies must grow in size with $\alpha = 0.8\pm0.4$ (cf. $\alpha = 1.27$ from \citealt{vanDokkum10}) to match the $z=0$ comparison sample mean. 
Metallicity gradients should therefore flatten with time as a natural consequence of inside-out growth, and the amount of size growth required is in rough agreement with direct measurements.

\section{Chemical Evolution Model}\label{sec:model}

As we have seen a first clear trend of flattening in the metallicity gradients for comparable systems with cosmic time, we seek now to develop a physically-motivated explanation
of the trend. Our purpose in this section is to explain the metallicity of a star-forming system (specifically a galaxy or a region within a galaxy) in terms of gas accretion, outflows, and star formation. The simplest such model is the ``closed box'' in which accretion and outflow rates are both zero. In this case the gas-phase metal mass fraction $Z$ is determined entirely by the yield $y$ and gas fraction $f_{\mathrm{gas}}$ as
\begin{equation}
Z = y \ln{f_{\mathrm{gas}}^{-1}}
\end{equation}
\citep{Schmidt63}.
This can be equivalently expressed in terms of the oxygen number abundance,
\begin{equation}
12+\log{O/H} = 12 + \log{\frac{y_0}{11.728}} + \log{(\ln{f_{\mathrm{gas}}^{-1}})}
\end{equation}
where $y_0$ is the oxygen yield and the factor 11.728 converts oxygen mass abundance to number abundance \citep{Lee06}. Throughout this paper we assume a value $y_0 = 0.0087$, appropriate for solar abundance ratios with $y=0.02$ \citep{Finlator08}.

\begin{figure}
\centerline{\includegraphics[width=0.5\textwidth]{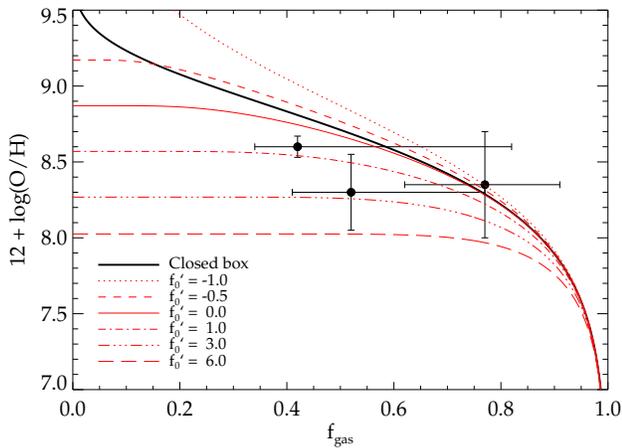}}
\caption{\label{fig:model} 
Gas phase metallicity as a function of gas fraction. We show predictions from our chemical evolution model as well as the closed box model. Data points correspond to measured values of the galaxies J0744, J1148, and J1206.
}
\end{figure}

We now construct a model which includes accretion and outflow of gas. The star-forming system which we model has the following properties: a gas mass $M_g$, stellar mass $M_*$, star formation rate SFR\,$=\dot{M_*}$, total mass of metals in the gas $M_Z$, and metallicity defined as $Z_g = M_Z/M_g$. We consider four factors which affect the metallicity at a given time:
\begin{enumerate}
\item {\em Inflowing gas.} The gas inflow rate is assumed to be a constant multiple $f_i$ of the star formation rate, $\dot{M_i} = f_i \dot{M_*}$. Inflowing gas is assumed to be metal-free ($Z=0$) corresponding to pristine gas accreted from the intergalactic medium.
\item{\em Transported gas.} In addition to inflow $\dot{M_i}$, we assume that some accreted gas is already enriched with a metallicity $Z=Z_g$ and accreted at a constant multiple of the star formation rate, $\dot{M_{en}}=f_{en} \dot{M_*}$. If the system considered is a region within a galaxy, then this term corresponds to the transportation of metal-enriched gas within the galaxy. If the system is a galaxy, this can be thought of as a ``galactic fountain'' term which describes the re-accretion of gas which has been expunged in outflows.
\item {\em Outflowing gas.} Outflows are a vital component in any reasonable model of galaxy chemical evolution. Outflows are observed ubiquitously in galaxies with star formation rates $\gsim 0.1$ \Msunyr kpc$^{-2}$ \citep{Heckman02} which is greatly exceeded by the lensed galaxies studied in this paper. The outflowing gas is assumed to have metallicity $Z=Z_g$ and a rate proportional to the SFR, $\dot{M_o} = f_o \dot{M_*}$. The ratio of outflow rate to SFR, $f_o$, is called the mass loading factor (MLF).
\item {\em Star formation.} Star formation is ultimately the source of all heavy elements. The amount of metals produced per unit time is defined as the yield multiplied by the star formation rate, $y \dot{M_*}$. Star formation additionally removes metals from the gas reservoir at a rate $Z \dot{M_*}$.
\end{enumerate}
We additionally assume that the gas mass remains constant. With these definitions, the metallicity is expressed as
\begin{equation}\label{eq:money}
\begin{split}
12+\log{O/H} = &12 + \log{\frac{y_0}{11.728}} - \log{(1+f_o')} \\
&+ \log{\left[ 1 - \exp{\left( -(1+f_o')\frac{1-f_{\mathrm{gas}}}{f_{\mathrm{gas}}} \right)} \right]}
\end{split}
\end{equation}
where we define the ``effective mass loading factor'' (EMLF) $f_o' = f_o - f_{en}$.
(A complete derivation of Equation~\ref{eq:money} is provided in Appendix~\ref{app:model}, along with an assessment of the assumptions made.)
Therefore, our model provides the gas phase metallicity as a function of EMLF and gas fraction. Equivalently, $f_o'$ can be determined from measurements of metallicity and gas fraction. We show the results of this model for various values of $f_o'$ in Figure~\ref{fig:model} along with measurements of the lensed galaxies.

\subsection{$f_o'$ as the Origin of Metallicity Gradients}\label{sec:origin}

The presence of metallicity gradients in local disk galaxies is typically attributed to variations in the mass loading factor with radius and/or inward radial migration of metal-enriched gas. Both of these physical effects can be expressed as a variation of $f_o'$ with radius, where negative metallicity gradients imply that $f_o'$ increases with radius. In Appendix~\ref{app:model} we derive the relation between gradients in metallicity and $f_o'$, and here we apply these findings to the lensed galaxies. Equation~\ref{eq:dz} shows that a gradient in $f_o'$ (expressed as $\Delta_R = \frac{d f_o'}{d R}$) naturally gives rise to a metallicity gradient $\Delta_Z = \frac{d}{d R} (\log{O/H})$.
For fiducial lensed galaxy values $f_{\mathrm{gas}} \simeq 0.5$ and $f_o' \simeq 2$, the relation is given by
$$\Delta_Z = -0.12 \Delta_R$$
dex. A gradient $\Delta_R = 0.5-2$ kpc$^{-1}$ can therefore explain the observed metallicity gradients of the isolated galaxies (Table~\ref{tab:gradients2}), and the merger J1038 requires $\Delta_R = -0.7$ kpc$^{-1}$ to explain the inverted gradient.

Let us now consider the evolution of such a gradient with time under the assumption that the gas fraction does not vary significantly with radius. In this case the time derivative $\frac{d \Delta_Z}{d t}$ is given by Equation~\ref{eq:ddzdt}.
An important qualitative feature of this equation is that $\frac{d \Delta_Z}{d t}$ is {\em negative} if $\frac{d \Delta_R}{d t} \geq 0$, in which case metallicity gradients will become stronger with time. This is easily understood in terms of Figure~\ref{fig:model}. Physically, we have assumed that the gas fraction is constant throughout a galaxy and declines monotonically with time. 
As $f_{gas}$ decreases, the difference in metallicity between two regions with different $f_o'$ will become larger and hence gradients will become more pronounced. However, it is unlikely that $\Delta_R$ increases with time. For example, taking a typical value $f_o'=2$ at a radius $R=1$ kpc, the eventual metallicity is only $12+\log{O/H} = 8.4$ or about half the solar value. Descendents of these lensed galaxies typically have super-solar metallicity at such small radii, implying that $f_o'$ must decrease at later times. The required decrease in mass loading factor can be quantified from the observational constraint that $\frac{d \Delta_Z}{d t} \lsim 0.05$ dex\,kpc$^{-1}$\,Gyr$^{-1}$ (Figure~\ref{fig:gradients5}). Using fiducial values for the lensed galaxies $f_{\mathrm{gas}} \simeq 0.5$, $f_o' \simeq 2$, $\dot{M_*} \simeq 100$ \Msunyr, $M_{\mathrm{gas}} \simeq 2\times10^{10}$ \Msun, and $\Delta_R \simeq 2$ kpc$^{-1}$, Equation~\ref{eq:ddzdt} implies that $\frac{d \Delta_R}{dt} \simeq -4$ kpc$^{-1}$\,Gyr$^{-1}$. Therefore the effective mass loading factor must decrease on a time scale $t_1 = \Delta_R\,(\frac{d \Delta_R}{dt})^{-1} \sim 500$ Myr, comparable to the characteristic age of  recent star formation in the lensed galaxies. 
This result suggests that a phase of strong starburst-driven outflows with high mass loading factors lasts only a few hundred Myr, which matches the duty cycle of strong star formation in LBGs inferred from statistical studies (e.g. \citealt{Stark09}).

In summary, we have used the chemical evolution model to show that a radial gradient in the effective mass loading factor $\Delta_R$ produces a metallicity gradient. The observed time evolution of metallicity gradients requires that $\Delta_R$ must decrease with time, and the characteristic starburst-driven outflow time scale from our model ($\sim 500$ Myr) is in rough agreement with the duty cycle of star formation inferred from statistical measurements of high redshift LBG demographics.

\section{Discussion}

We now discuss what can be learned physically about the mass assembly in $z\simeq2$ galaxies by applying the model discussed in \S\ref{sec:model} to measurements of the galaxies in our sample.

\subsection{Mass Loading Factor}\label{sec:mlf}

The mass loading factor, defined as the ratio $f_o$ of mass outflow rate to star formation rate, is an important ingredient of galaxy formation models. Our simple chemical evolution model can be used to infer $f_o$ from measurements of metallicity and gas fraction via Equation~\ref{eq:money}. We apply this equation to the integrated values of the lensed galaxies (Table~\ref{tab:mass}) and list the resulting effective mass loading factors $f_o'$ in Table~\ref{tab:model}. Assuming that accreted gas has an average metallicity significantly lower than that of the galaxies, we set $f_{en} \approx 0$ and equate $f_o' = f_o$ (see discussion in \S\ref{sec:validity}). Therefore, the model implies a modest mass loading factor $f_o<2.3$ for all three galaxies studied in detail here.

In the case of J1148, we can estimate the true mass loading factor from direct measurements of the outflow column density. \cite{Quider09} measure a column density $\log N_{\mathrm{Si{\sc II}}} = 16$ cm$^{-2}$ in the outflowing gas. The Si{\sc ii}/Si{\sc iv} ratio is $\simeq 12$ suggesting that most of the outflowing mass is in neutral Hydrogen gas traced by Si{\sc ii}. Motivated by our chemical evolution model and the results discussed later in this section (cf. Figure~\ref{fig:emlf_r}), we assume that most of the outflow occurs at large radius with metallicity $12+\log{O/H} \sim 8.1$ (Figure~\ref{fig:gradients}) or [O/H]~$\sim -0.6$. Since O and Si are both $\alpha$ elements generated by the same nucleosynthetic processes, we assume that [Si/H]~$\sim -0.6$ as well. Indeed, \cite{Pettini02} show that variation in $\alpha$ element abundances is $< 0.2$ dex in the ISM of the Lyman break galaxy cB58. Adopting a solar abundance ratio [O/H]$_{\odot} = -4.44$ \citep{Pettini02} gives an outflow column density $\log N_H = 21.0$ cm$^{-2}$. This corresponds to a total mass column density (including Helium) $\Sigma_{M_o} = 1.1 \times 10^7$ \Msun kpc$^{-2}$. We can crudely estimate the total mass assuming that outflowing gas uniformly fills a sphere of radius $R$:
$$M_o = \frac{4}{3} \pi R^2 \Sigma_o \sim 3 \times 10^9 ~\mathrm{\Msun}$$
for $R = \Delta v M_* / \dot{M_*} \sim 8$ kpc, outflow velocity $\Delta v = 200$ \kms \citep{Quider09}, and timescale $M_* / \dot{M_*} = 40$ Myr (Table~\ref{tab:model}). The time-average mass loading factor is then given by $f_o = M_o / M_* \sim 0.4$. Uncertainty in $f_o$ due to the assumed geometry is approximately an order of magnitude, such that we can confidently assert that $f_o \leq 5$ in reasonable agreement with the value derived from our chemical evolution model.

We now turn to spatial variations in the mass loading factor, and radial gradients in particular. 
Gas fractions in the lensed galaxies have almost no variation with radius, and therefore we can apply the formalism developed in \S\ref{sec:origin} to estimate $\Delta_R$, $\frac{d \Delta_R}{d t}$, and the characteristic timescale $t_1$. The results are listed in Table~\ref{tab:model}. We also give the mass assembly timescale $M_* / \dot{M_*}$.
We can compare the simple analytic treatment of \S\ref{sec:origin} with variations in $f_o'$ inferred directly from spatially-resolved measurements via Equation~\ref{eq:money}. We use the measured radial profiles of gas fraction and metallicity to compute $f_o'$, and show the results in Figure~\ref{fig:emlf_r}. As expected, $f_o'$ increases at large radii. The analytic approximation $\Delta_R$ is in reasonable agreement with a linear fit to the data, although the measured radial gradient of $f_o'$ is poorly represented by a straight line. 
Metallicity gradients observed in the lensed galaxies are therefore likely to originate from radial variations of the effective mass loading factor.

\subsection{Inflow Rate}

Under the equilibrium condition that the gas mass remains constant, the inflow rate is directly related to star formation rate and mass loading factor via Equation~\ref{eq:fi}. When considering a galaxy as a whole, $f_{en}$ is negligible and thus $f_i = 1+f_o$, or $\dot{M_i} = (1+f_o) \dot{M_*}$. We calculate $\dot{M_i}$ using the relevant values in Tables~\ref{tab:mass} and \ref{tab:model} and list the results in Table~\ref{tab:model}. Infall rates range from $1.2-2.0\times$ the star formation rate.

\subsection{Radial Gas Transport}

\begin{figure*}
\includegraphics[width=0.38\textwidth]{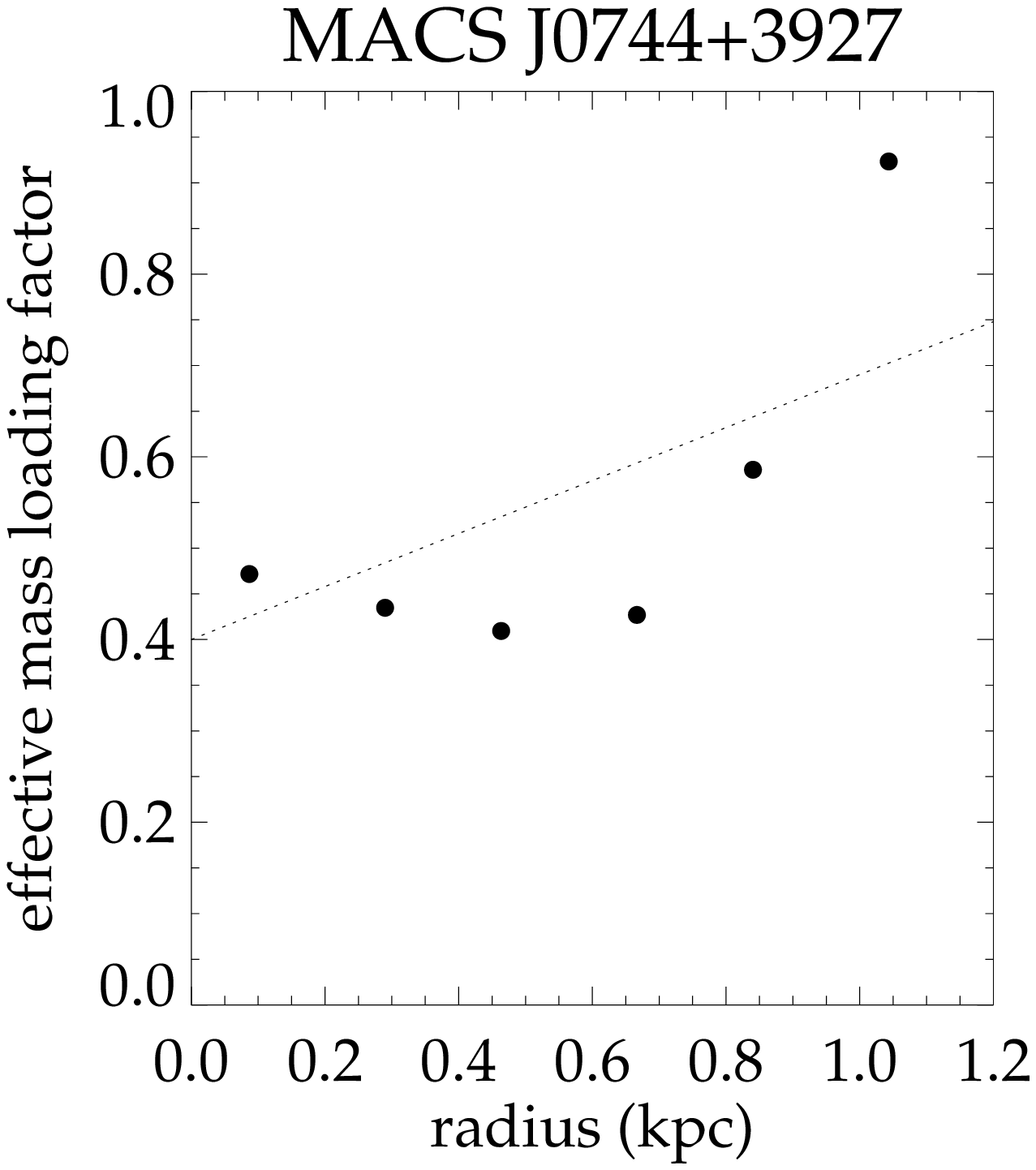}
\hspace{-.07\textwidth}
\includegraphics[width=0.38\textwidth]{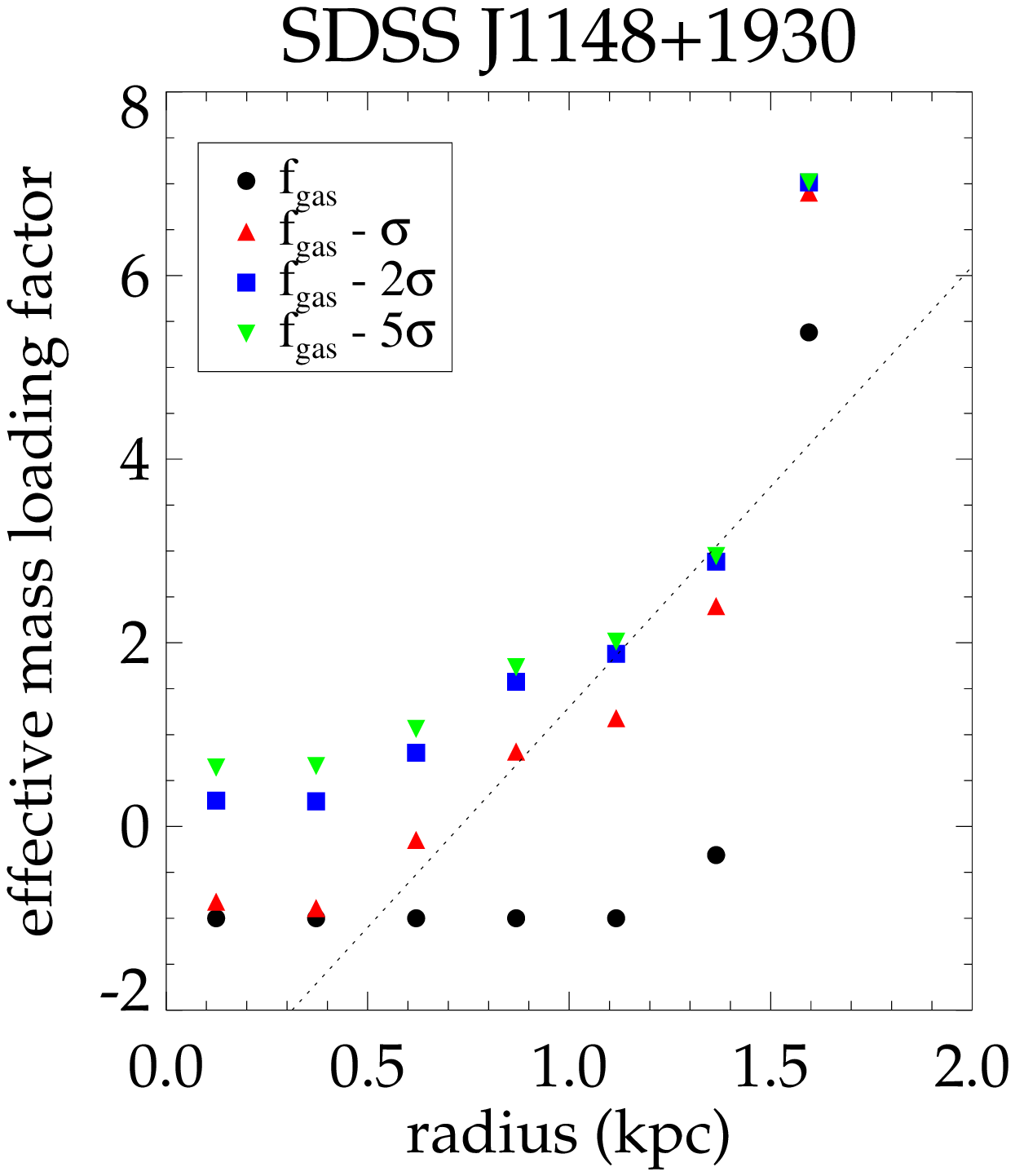}
\hspace{-.07\textwidth}
\includegraphics[width=0.38\textwidth]{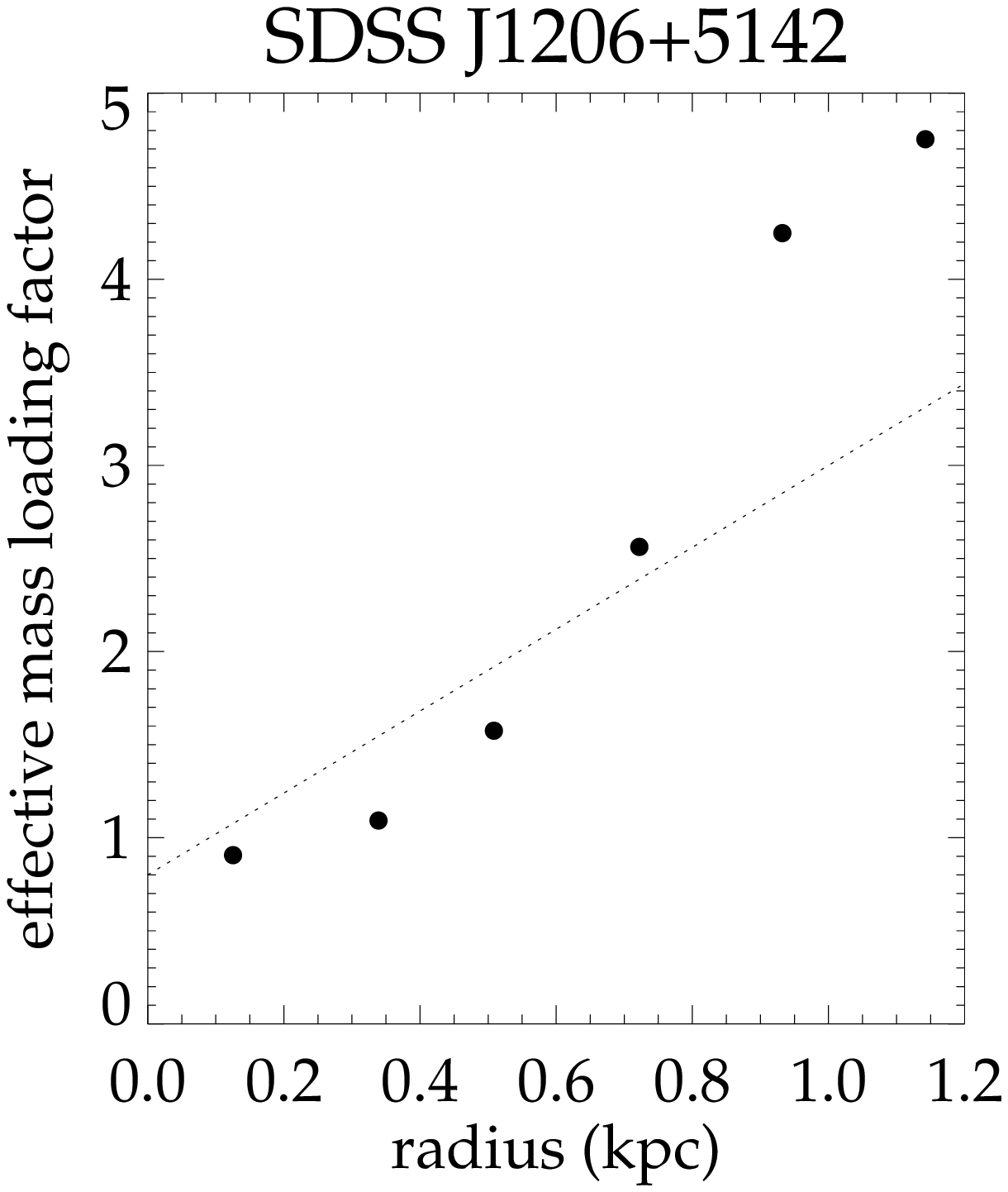} \\
\caption{\label{fig:emlf_r} 
Effective mass loading factor $f_o'$ vs. radius. $f_o'$ is calculated from Equation~\ref{eq:money} and is shown to increase with radius. The measured gas fraction and metallicity of J1148 is unphysical at $R<1$ kpc in the context of our chemical evolution model, and so we also show the results of perturbing the gas fraction by multiples of the uncertainty $\sigma(f_{\mathrm{gas}})=0.15$. We note that the overall shape and normalization of $f_o'(R)$ is relatively constant over a range $\Delta f_{\mathrm{gas}} = 4\sigma = 0.6$; clearly Equation~\ref{eq:money} is relatively insensitive to gas fraction in this regime. Dotted lines show the result from the simple analytic approximation described in \S\ref{sec:origin} (see discussion in the text). For J1148 the analytic prediction applies to the case $f_{\mathrm{gas}}-\sigma$.
}
\end{figure*}

We have suggested that radial metallicity gradients are caused by gradients in the effective mass loading factor. Negative gradients can be caused by increasing mass loading factor with radius ($\frac{d f_o}{dR} > 0$), by radial transport of enriched gas ($\frac{d f_{en}}{dR}<0$), or both. The degeneracy between $f_o$ and $f_{en}$ captured by Equation~\ref{eq:emlf} essentially means that the chemical evolution model cannot distinguish the contributions of outflow and gas transport. However, in cases where $f_o'<0$, gas transport is necessarily present. Figure~\ref{fig:emlf_r} shows that the inner regions of J1148 are inferred to have $f_o'<0$ with $\sim 2\sigma$ significance from uncertainty in the gas fraction, although the significance is marginal when considering a number of other assumptions made (e.g. the Schmidt relation, constant mass-to-light ratio, and the model itself). Nonetheless the high metallicity and high inferred gas fraction suggest that gas in the inner regions of J1148 has likely been enriched by previous star formation episodes at larger radii. This observation, combined with the clumpy structure and low $Q\lsim1$ of J1148, are in qualitative agreement with the model of \cite{Dekel09} in which gravitational instability leads to the formation of giant clumps which migrate into the galactic center in $\lsim500$ Myr.

We reiterate that this result depends crucially on the gas fraction, which we have estimated using the Kennicutt-Schmidt law. Future measurements of the gas density e.g. with ALMA are needed for direct verification.

\subsection{Positive Metallicity Gradients}

In the local universe, disk galaxies almost invariably show negative metallicity gradients. Only a few positive gradients have been measured, all of which are relatively shallow ($<0.05$ dex/kpc) and found exclusively in merging systems (see Figure~\ref{fig:gradients2}). In contrast, {\it positive} metallicity gradients have been claimed for some high redshift galaxies, including many with no signs of recent interaction or merging \citep{Cresci10,Queyrel12}. In the lensed sample we find that all three isolated galaxies have negative gradients, while the merging system J1038 has a positive gradient. What could explain these claimed positive gradients at high redshifts? \cite{Cresci10} and \cite{Queyrel12} have suggested they could be caused by high accretion rates of metal-poor gas which occur only at high redshift or during mergers.
Even so it is difficult to form a positive gradient since the fraction of gas converted into stars per unit time must {\em increase} with radius, while the dynamical time scale generally decreases with radius.
The corollary is that a higher fraction of metal-enriched gas must be lost to outflows at small radii, or in terms of our chemical evolution model, $f_o$ must decrease with radius. While gaseous infall is notoriously difficult to detect at high redshift, outflows are relatively easy to detect and in principle $f_o$ can be estimated as a function of radius following the method of \S\ref{sec:mlf}. Current planned and ongoing observations will directly address the variation of $f_o$ with radius in high redshift galaxies and determine whether this is a viable mechanism of producing positive metallicity gradients (see \S\ref{sec:future}).

Alternatively, it is possible that many of the positive (and negative) gradients reported at high redshift are a result of systematic errors rather than actual metallicity gradients. In this paper we have primarily used the ratio of \Nii/\Ha\, as a proxy for metallicity, however this alone is not convincing because of the issues discussed in \S\ref{sec:metallicity}. For example, \cite{Wright10} describe a galaxy with strong radial line ratio gradients which mimic a negative metallicity gradient in \Nii/\Ha\, and a positive gradient in \Oiii/\Hb. Using the diagnostic diagram of Figure~\ref{fig:bpt}, however, \cite{Wright10} show that these are caused by a central AGN surrounded by an extended star-forming disk rather than a metallicity gradient.
In the local universe, \cite{Westmoquette12} find that \Nii/\Ha\, increases with radius in 7/18 galaxies in their volume-limited sample and show, via measurements of [S{\sc ii}] and [O{\sc i}] emission, that these are caused by an ionization parameter gradient. The most likely explanation is an increasing contribution of shocks at large radius rather than a positive metallicity gradient. Therefore it is known that the line ratio signatures of both positive and negative metallicity gradients are mimicked by AGN and shocks, and we caution that gradients based on strong line ratios are not reliable unless confirmed with multiple appropriate diagnostics (e.g. Figures~\ref{fig:bpt} and \ref{fig:gradients_compare}).

\begin{table*}
\begin{tabular}{lccccccccc}
Name        &  12+log\,O/H  &  $f_o'$  &  $\Delta_R$   &  $\frac{d \Delta_R}{d t}$  &  $t_1$  &  $M_* / \dot{M_*}$   &  model $\dot{M_i}$  \\
            &                  &       &  (kpc$^{-1}$) &  (kpc$^{-1}$\,Gyr$^{-1}$)   &  (Myr)  &  (Myr)              &   (\Msunyr)         \\
\hline
J0744       &  8.65  &  $0.4^{+0.2}_{-1.4}$ &  0.29        &  -0.38  &  760  &  470  &  8    \\
J1148       &  8.40  &  $-1.0^{+2.2}_{-0}$  &  4.8         &  -25    &  110  &   40  &  250  \\
J1206       &  8.35  &  $2.1^{+0.2}_{-3.1}$ &  2.2         &  -4.8   &  450  &  190  &  140  \\
\end{tabular}
\caption{\label{tab:model} Properties inferred from the chemical evolution model. Effective mass outflow rates and uncertainties correspond to the formal uncertainties in gas fraction for the tabulated value of $12+\log{O/H}$. $\Delta_R$, its derivative, and $t_1$ are calculated from $f_o'$ except in the case of J1148 where we use $f_o' + \sigma = 1.2$.}
\end{table*}

\subsection{Future Work}\label{sec:future}

The study of galaxy formation through time evolution of metallicity gradients is still in its infancy. Here we briefly discuss three areas of future work that will improve our understanding: the mass loading factor, sample sizes and redshift range, and gas fractions.

We have shown from simple chemical evolution arguments that metallicity gradients likely originate from radial gradients in the mass loading factor $f_o$. Figure~\ref{fig:emlf_r} suggests that this quantity is several times higher at large radius ($R\gsim1$ kpc for the lensed galaxies) than at the center. Furthermore, we have shown that a radially decreasing $f_o$ can create the positive metallicity gradients observed at high redshift but for which there are no local analogs.
While difficult to measure accurately, $f_o$ can be estimated from rest-UV absorption lines as described in \S\ref{sec:mlf} and spatially resolved measurements can constrain its variation with radius. We are actively pursuing these measurements in lensed galaxies at $z\sim2$ via an approved program with the Keck II/ESI integral field unit.

The sample of high redshift galaxies with robust metallicity gradient measurements remains very small. Although data exist for more than 30 galaxies at $z>1$, the majority lack the data needed to distinguish true metallicity gradients from the signatures of shocks and AGN. Furthermore, different samples reported at $z>1$ (\citealt{Cresci10, Yuan11, Queyrel12}; this work) are discrepant and warrant further investigation. Additionally the range of redshifts $0.1 < z < 1$ has not yet been explored, and our data suggest that gradients should decrease in magnitude by a factor of $\simeq2$ in this time (e.g. Figure~\ref{fig:gradients5}).  
It is important to realize that highly multiplexed optical spectrographs can efficiently measure resolved emission line fluxes to large galactocentric radius in statistically significant samples of galaxies out to $z\lsim1.7$ \citep{Miller11,Miller12}. This is an attractive means of procuring a large number of metallicity gradient measurements at intermediate redshifts $z\lsim1$, especially in tandem with multi-object near infrared spectrographs to detect \Ha\, and \Nii\, at $z\gsim0.5$, and we will exploit these capabilities in future work.

Finally, results from chemical evolution modelling are sensitive to the gas fraction which is difficult to measure at high redshift. In the present work we have estimated gas masses by inverting the Kennicutt-Schmidt relation, introducing a considerable degree of uncertainty. Gas mass is difficult to determine even with direct measurements and typically relies on an uncertain factor $X_{CO}$ to convert a CO line luminosity to gas mass. With the advent of ALMA it is now possible to resolve the molecular gas emission in typical star forming galaxies at high redshift, and calibrate gas mass-to-luminosity ratios from independent dynamical measurements (e.g. \citealt{Daddi10, Stark08}). With such advances, we may be able to confidently determine resolved gas fractions in the near future.

\section{Summary}

This paper presents robust spatially resolved metallicity measurements of four lensed galaxies at $z=2-2.4$ based on the strong emission lines \Ha, \Nii, \Oiii, and \Hb. The combination of gravitational lensing and adaptive optics provides a source plane resolution of up to $300-600$ parsecs in each galaxy. Three targets are isolated rotating galaxies, and one is a merger of at least two systems with mass ratio $(6\pm3):1$. All three rotating galaxies have negative radial metallicity gradients inferred from the \Nii/\Ha\, emission line ratio and confirmed with \Oiii\, and \Hb\, measurements. The merging system has a positive gradient in the UV-bright source, also confirmed with multiple emission line ratios.

Metallicity gradients measured at $z>2$ are compared with galaxies at lower redshift selected to occupy equivalent dark matter halos. On average, gradients in the rotating galaxies must flatten by a factor of $2.6 \pm 0.9$ between $z=2.2$ and $z=0$. This factor is in rough agreement with size evolution measured for more massive galaxies by \cite{vanDokkum10}, hence radial inside-out growth can account for the variation in metallicity gradients with redshift.

We develop a simple model of chemical evolution to explain the emerging body of data. From a combination of spatially resolved gas fraction and metallicity, the model implies that metallicity gradients are caused by radial variation in the effective mass loading factor. The time-averaged mass loading factor is inferred to be $\lsim 2$.
Finally, inward radial gas transport is required to explain the high metallicity and gas fraction of J1148, and is likely present in the other galaxies although we cannot determine this directly from the model.

The study of metallicity gradients at different redshifts is still relatively new and robust sample sizes are limited. Future work addressing direct measurements of the molecular gas content, metallicity gradients at intermediate redshifts ($0.2 \lsim z \lsim 1$), and direct, spatially resolved measurements of the mass loading factor are planned or ongoing. These new observations will confirm (or refute) the results of this paper and will place additional constraints on the formation of galaxies from the origin and evolution of metallicity gradients.

\section*{Acknowledgments}
We thank the staff of Keck Observatory for their tireless support of these observations.
TAJ acknowledges stimulating discussions with Shy Genel, Fabio Bresolin, Judy Cohen, Wal Sargent, Romeel Dav{\' e}, and Hai Fu.
JR is supported by the Marie Curie Career Integration Grant 294074. EJ acknowledges support from the NASA Postdoctoral Program and CNRS.

Most of the data presented herein were obtained at the W.M. Keck Observatory, which is operated as a scientific partnership among the California Institute of Technology, the University of California and the National Aeronautics and Space Administration. The Observatory was made possible by the generous financial support of the W.M. Keck Foundation.
The authors wish to recognize and acknowledge the very significant cultural role and reverence that the summit of Mauna Kea has always had within the indigenous Hawaiian community.  We are most fortunate to have the opportunity to conduct observations from this mountain.

\appendix
\section{Chemical Evolution Model}\label{app:model}

In this Appendix we provide the mathematical details of the chemical evolution model discussed in the text. All notation and definitions are the same as in \S\ref{sec:model}. To begin, the combination of inflowing gas, transported gas, outflowing gas, and star formation as described in \S\ref{sec:model} gives a rate of change of metal mass
\begin{equation}\label{eq:dmzdt}
\frac{d M_z}{d t} = y \dot{M_*} - Z_g \dot{M_*} - Z_g \dot{M_o} + Z_g \dot{M_{en}}.
\end{equation}
The gas mass is assumed to be constant, i.e.
\begin{equation}\label{eq:mgas}
\dot{M_g} = \dot{M_i} + \dot{M_{en}} - \dot{M_o} - \dot{M_*} = 0
\end{equation}
or equivalently
\begin{equation}\label{eq:fi}
f_i + f_{en} - f_o - 1 = 0.
\end{equation}
Equation~\ref{eq:dmzdt} therefore simplifies to
\begin{equation}\label{eq:dMz}
\frac{d M_z}{d t} = y \dot{M_*} - Z_g \dot{M_*} - Z_g f_o'\dot{M_*}.
\end{equation}
where we define the ``effective mass loading factor'' (EMLF) $f_o'$ as
\begin{equation}\label{eq:emlf}
f_o' = f_o - f_{en}.
\end{equation}
The rate of change of metallicity is
\begin{equation}
\frac{d Z_g}{d t} = \frac{d}{dt} \frac{M_Z}{M_g} = \frac{1}{M_g} \frac{d M_z}{d t} 
= \frac{\dot{M_*}}{M_g} (y - Z_g - Z_g f_o')
\end{equation}
from Equation~\ref{eq:dMz}.
Integrating with respect to time gives
\begin{equation}
- \frac{1}{1+f_o'} \ln{(y - Z_g - Z_g f_o')} + \frac{1}{1+f_o'} \ln{(y)} = \frac{M_*}{M_g} 
= \frac{1-f_{\mathrm{gas}}}{f_{\mathrm{gas}}}
\end{equation}
where we introduce the gas fraction defined in Equation~\ref{eq:fgas} and apply the boundary conditions $Z_g=0$ and $M_*=0$ at time $t=0$. With some algebra we obtain the desired expression for metallicity,
\begin{equation}\label{eq:money1}
Z_g = \frac{y}{1+f_o'} \left[ 1 - \exp{\left( -(1+f_o')\frac{1-f_{\mathrm{gas}}}{f_{\mathrm{gas}}} \right)} \right].
\end{equation}
While Equation~\ref{eq:money1} contains all of the information needed, it is more convenient to express the metallicity in terms of 12+log(O/H). The equation then becomes
\begin{equation}\label{eq:money_app}
12+\log{O/H} = 12 + \log{\frac{y_0}{11.728}} - \log{(1+f_o')} 
+ \log{\left[ 1 - \exp{\left( -(1+f_o')\frac{1-f_{\mathrm{gas}}}{f_{\mathrm{gas}}} \right)} \right]}
\end{equation}
as given in the text.

We now derive the relation between gradients in metallicity and $f_o'$ and their evolution with time. 
From Equation~\ref{eq:money_app}, metallicity varies with $f_o'$ as
\begin{equation}
\frac{d}{d f_o'} (\log{O/H}) = \frac{1 + x - \exp{(x)}}{\ln(10) (1+f_o') [\exp{(x)}-1]}
\end{equation}
where $x = (1+f_o')(1-f_{\mathrm{gas}})/f_{\mathrm{gas}}$. This gives rise to a metallicity gradient of magnitude
\begin{equation}\label{eq:dz}
\Delta_Z = \frac{d}{d R} (\log{O/H}) = \Delta_R \frac{d}{d f_o'} (\log{O/H})
\end{equation}
where we define the $f_o'$ gradient $\Delta_R = \frac{d f_o'}{d R}$. A radial gradient in $f_o'$ therefore naturally results in a non-zero metallicity gradient.
It is instructive to examine the resulting evolution of metallicity gradients with time. Noting that the gas fraction was found to be approximately constant with radius in the lensed galaxies (\S\ref{sec:fgas}), we make the simplifying assumption that $\frac{d f_{\mathrm{gas}}}{d t} = 0$. In this case the rate of change of a metallicity gradient is given by
\begin{equation}\label{eq:ddzdt}
\begin{split}
\frac{d \Delta_Z}{d t} &= \frac{d}{d t} \Delta_R \frac{d}{d f_o'} (\log{O/H}) \\
&= \Delta_R \frac{d^2}{dt\,df_o'} (\log{O/H}) + \frac{d \Delta_R}{d t} \frac{d}{d f_o'} (\log{O/H}) \\
&= \Delta_R \frac{1}{\ln(10)} \frac{\dot{M_*}}{M_{\mathrm{gas}}} \frac{(1-x) \exp{(x)} - 1}{[\exp{(x)} - 1]^2} + \frac{d \Delta_R}{d t} \frac{1 + x - \exp{(x)}}{\ln(10) (1+f_o') [\exp{(x)}-1]}.
\end{split}
\end{equation}

\subsection{Validity of the Model}\label{sec:validity}

We now discuss important assumptions of the chemical evolution model, their validity, and quantify the implications for interpreting data using this model. There are several assumptions with which we are concerned.

Perhaps the most important assumption is that total gas mass is constant (Equation~\ref{eq:mgas}). Given that gas accretion and outflows are undoubtedly present, we expect the gas mass to fluctuate with time. However, we also expect the gas reservoir to be in equilibrium between accretion, star formation rate, and outflows. Cosmological simulations which incorporate outflows indeed show that the observed mass-metallicity relation and its dependence on star formation rate can indeed be explained in terms of such an equilibrium (e.g. \citealt{Dave11}). In these simulations, departure from this equilibrium gives rise to scatter about the observed mass-metallicity-SFR relation, which is measured to be only 0.05 dex in metallicity among local star-forming galaxies \citep{Mannucci10}. Furthermore, if the gas mass remains constant, then the expected growth in stellar mass (see \S\ref{sec:z_evo}) results in gas fractions $f_{\mathrm{gas}} \simeq 0.03-0.15$ at $z=0$ for the lensed galaxies. This is in good agreement with direct measurements of gas fractions in local galaxies with appropriate stellar masses $\log{M_*} \sim 10.7$ (e.g. \citealt{Young91}). We therefore expect the assumption that $\dot{M_g} = 0$ to be approximately valid, resulting in an uncertainty of only $\simeq 0.05$ dex in 12+log(O/H).

Another critical assumption is that gas infall and outflow rates ($f_i$, $f_o$, $f_{en}$) are a constant fraction of the SFR. This is clearly not physically true; these rates should therefore be interpreted as a {\em time-averaged} value over the formation history of the galaxy rather than instantaneous rates. Observationally it is well known that nearly all galaxies observed at high redshift drive outflows of gas, while outflows are much less common in their descendants at $z\simeq0$. Therefore $f_o$ must decrease with time, while $f_i$ and/or $f_{em}$ must decrease accordingly to maintain equilibrium. Therefore we expect that model-inferred $f_i$, $f_o$, and $f_{en}$ correspond to time-averaged values and are generally larger than the instantaneous rates.

We have assumed in the model that accreted gas is divided into two components with metallicity $Z=0$ and $Z=Z_g$. This is entirely for ease of interpretation in the context of this paper. In reality accreted gas will most likely have an average metallicity within the range $0<Z<Z_g$; the values of $f_i$ and $f_{em}$ can be adjusted to reach the desired value. However for our analysis it is more convenient to think of the accreted gas as a pristine cosmological component and an enriched component, as formulated. If the enriched component arises from metal transport within a galaxy, then the assumption $Z=Z_g$ is equivalent to the condition that metallicity is a continuous function of position. This is necessarily true. The assumption that infalling gas has $Z=0$, however, is not true. The infalling gas will have a metallicity $Z_i$, introducing an additional term $+Z_i \dot{M_i}$ to the right side of Equation~\ref{eq:dMz}. Functionally this can be though of as a modification to the yield, such that the effective yield is
\begin{equation}
y_{eff} = y + f_i Z_i.
\end{equation}
The IGM oxygen abundance is measured to be [O/H]~$=-2.8$ at $z=2.5$ \citep{Simcoe04} corresponding to $Z_i = 0.0015 \, Z_{\odot} = 3 \times 10^{-5}$. For $y_0 = 0.0087$, the effect of IGM metallicity is negligible ($<10$\% of $y_0$) for $f_i < 30$. For an infall rate $f_i = 7$ corresponding to the highest acceptable value for the lensed galaxies (see Figure~\ref{fig:model}), the effective yield differs from $y_0$ by only 2.5\%. For the redshifts and metallicities of interest in this paper, the assumption that accreted gas has $Z=0$ therefore introduces a negligible systematic error of $\leq2.5$\% or $\leq 0.01$ dex in metallicity.

In summary, the assumptions used in our chemical evolution model are in reasonable agreement with the best current data and understanding of galaxy formation and of the IGM. The systematic uncertainties introduced by our assumptions are expected to be negligible (e.g. $\lsim 0.05$ dex in metallicity). We caution that the inflow and outflow rates must be treated as time-averaged rather than instantaneous values. Values of $f_o$ and $f_i$ inferred from the model are expected to overestimate the current instantaneous rates.

\footnotesize

\bibliography{deimos}{}
\bibliographystyle{apj}
\end{document}